\documentclass{aa}
\usepackage{graphicx}

\newcommand{\diff}{\mbox{${\rm d}$}}

\newcommand{\ub}{\mbox{$U\!-\!B$}}
\newcommand{\bv}{\mbox{$B\!-\!V$}}
\newcommand{\vi}{\mbox{$V\!-\!I$}}
\newcommand{\vr}{\mbox{$V\!-\!R$}}
\newcommand{\vk}{\mbox{$V\!-\!K$}}
\newcommand{\jh}{\mbox{$J\!-\!H$}}
\newcommand{\jk}{\mbox{$J\!-\!K$}}
\newcommand{\hk}{\mbox{$H\!-\!K$}}
\newcommand{\mv}{\mbox{$M_{V}$}}

\newcommand{\mk}{\mbox{$M_{K}$}}

\newcommand{\feh}{\mbox{\rm [{\rm Fe}/{\rm H}]}}
\newcommand{\mh}{\mbox{\rm [{\rm M}/{\rm H}]}}
\newcommand{\Msun}{\mbox{$M_{\odot}$}}
\newcommand{\Teff}{\mbox{$T_{\rm eff}$}}

\newcommand{\comment}[1]{}
\newcommand{\beq}{\begin{equation}}
\newcommand{\eeq}{\end{equation}}
\newcommand{\beqa}{\begin{eqnarray}}
\newcommand{\eeqa}{\end{eqnarray}}
        \def\smallskip{\vskip 2pt}

\begin{document}

\title{Theoretical isochrones in several photometric systems} 
\subtitle{I. Johnson-Cousins-Glass, HST/WFPC2, HST/NICMOS, 
Washington, and ESO Imaging Survey filter sets}

\author{L. Girardi\inst{1,2}, G. Bertelli\inst{3,4}, 
A. Bressan\inst{4}, C. Chiosi\inst{1}, \\ M.A.T. Groenewegen\inst{5,6}, 
P. Marigo\inst{1}, B. Salasnich\inst{1}, A. Weiss\inst{7}}
\institute{
 Dipartimento di Astronomia, Universit\`a di Padova,
	Vicolo dell'Osservatorio 2, I-35122 Padova, Italy \and
 Osservatorio Astronomico di Trieste, 
	Via Tiepolo 11, I-34131 Trieste, Italy \and
 Istituto di Astrofisica Spaziale, CNR, Via del Fosso del Cavaliere, 
	I-00133 Roma, Italy \and
 Osservatorio Astronomico di Padova, 
	Vicolo dell'Osservatorio 5, I-35122 Padova, Italy \and
 PACS ICC-team, Instituut voor Sterrenkunde,
	Celestijnenlaan 200B, B-3001 Heverlee, Belgium \and
 European Southern Observatory, Karl-Schwarzschild-Str.\ 
	2, D-85740 Garching bei M\"unchen, Germany  \and
 Max-Planck-Institut f\"ur Astrophysik, Karl-Schwarzschild-Str.\ 
	1, D-85740 Garching bei M\"unchen, Germany 
}

\offprints{L\'eo Girardi \\ e-mail: Lgirardi@ts.astro.it} 

\date{Received 05-11-01 / Accepted 19-04-02}

\abstract{
We provide tables of theoretical isochrones in several 
photometric systems. To this aim, the following steps
are followed:
(1) First, we re-write the formalism for converting 
synthetic stellar spectra into tables of bolometric 
corrections. The resulting formulas can be  
applied to any photometric system, provided that 
the zero-points are specified by means of either ABmag, STmag, 
VEGAmag, or a standard star system that includes well-known 
spectrophotometric standards. Interstellar absorption can be 
considered in a self-consistent way. 
(2) We assemble an extended and 
updated library of stellar intrinsic spectra. It is mostly
based on ``non-overshooting'' ATLAS9 models, suitably extended
to both low and high effective temperatures. This offers an
excellent coverage of the parameter space of \Teff, $\log g$, 
and \mh. We briefly discuss the main uncertainties and points 
still deserving more improvement.
(3) From the spectral library, we derive tables of bolometric 
corrections for Johnson-Cousins-Glass, HST/WFPC2, HST/NICMOS, 
Washington, and ESO Imaging Survey systems (this latter
consisting on the WFI, EMMI, and SOFI filter sets). 
(4) These tables are used to convert several sets of Padova
isochrones into the corresponding absolute magnitudes and 
colours, thus providing a useful database for several astrophysical
applications. All data files are made available in electronic 
form.
}

\authorrunning{Girardi et al.}
\titlerunning{Isochrones in several photometric systems}
\maketitle

\section{Introduction}
\label{intro}

One of the primary aims of stellar evolution theory is that of
explaining the photometric data -- eg.\ colour-magnitude diagrams 
(CMDs), luminosity functions, colour histograms -- of resolved 
stellar populations. To allow whatever comparison between theory and 
data, the basic output of stellar models -- the surface 
luminosity $L$ and effective temperature \Teff\ -- must be first
converted into the observable quantities, i.e.\ magnitudes and 
colours. This conversion is performed by means of
bolometric corrections (BC) and \Teff-colour relations, and 
later by considering the proper distance, absorption and 
reddening of the observed population, and the photometric errors.

Determining BCs and \Teff-colour relations 
is indeed one of the most basic tasks in stellar astrophysics.
Empirical determinations (see e.g.\ the compilations by
Schmidt-Kaler 1982, Flower 1996, and Alonso et al.\ 
1999ab) involve a great observational effort, and
are obviously the most reliable if compared to purely
theoretical determinations. However, since the empirical calibrations
are mostly based on 
nearby stars, only a limited region in the space of stellar 
parameters (\Teff, $\log g$, and metallicity \mh) can be covered by 
observations. This contrasts with our
present-day capabilities of getting resolved photometry of populations
that are certainly very different from the local one, like those in dwarf
galaxies and in the Bulge.  For instance, present empirical relations
do not include young-metal poor populations, or super-metal rich stars,
which are likely present in the resolved galaxies of the Local Group.

Another important limitation of empirical relations is that they are
usually available just for a small set of filters or photometric
systems, like the popular Johnson-Cousins-Glass one. Again, this is in
contrast with the rapid diffusion in the use of specific filter sets,
for which no empirical relation is yet available, though large
databases are already being collected. Just to mention a few relevant
examples, new filter sets have been adopted in the Hubble
Space Telescope (HST) Wide Field Planetary Camera 2 (WFPC2), in the
European Southern Observatory (ESO) Wide Field Imager (WFI), and in
the Hipparcos mission. Moreover, brand-new photometric systems
have been designed for the Sloan Digital Sky Survey (SDSS), and will
be adopted in the future GAIA mission. An impressive amount of data
will be provided by these instruments in the coming years, and much of
it will soon become of public access. Then, it would be highly desirable
to have the capability of converting stellar models to these many
new photometric systems, bypassing for a moment the time-consuming
procedure of empirical calibration.

With this target in mind, we have undertaken a project aiming at providing
theoretical BCs and colour transformations for {\em any} broad-band
photometric system, and many intermediate-band systems as well.
Actually, the project starts with the work by
Bertelli et al.\ (1994), who presented a large database of theoretical
isochrones and converted them to the Johnson-Cousins-Glass $UBVRIJHK$ 
system. The transformations were primarily based on Kurucz (1993) 
synthetic atmosphere models, suitably extended in the intervals
of lower and higher effective temperatures. In Chiosi et al.\ (1997), 
the same theoretical isochrones are converted to the WFPC2 photometric
system, and the attention is paid to the features that 
isochrones and single stellar populations present in ultraviolet (UV)
colours. In particular, they address the question whether the 
variation of UV colours of elliptical galaxies as a function of 
red-shift presents signatures from which one can infer the age and type
of the source emitting the UV flux. Later on, Salasnich et al.\ (2000)
present new isochrones for $\alpha$-enhanced chemical mixtures,
for both $UBVRIJHK$ and WFPC2 photometric systems.

The present work is a natural follow-up of this project, in which,
besides extending the number of available photometric systems,
we aim at updating and improving the database of stellar spectra
which is at the basis of the complete procedure.

The plan of this paper is as follows:
Sect.~\ref{sec_synphot} details the adopted formalism. 
In Sect.~\ref{sec_spectra}, we describe the stellar spectral
library in use.  Sect.~\ref{sec_systems} gives the basic
information about the several photometric systems under consideration.
The resulting tables of bolometric corrections are then applied to a
large database of stellar isochrones, which are already described in 
published papers and briefly recalled in Sect.~\ref{sec_tracks}.
Sect.~\ref{sec_conclu} illustrates some main properties of the
derived isochrones, and describes the retrieval of data in electronic
form.
 
\section{Synthetic photometry}
\label{sec_synphot}

By synthetic photometry we mean the derivation of photometric 
quantities based on stellar intrinsic (and mostly theoretical) 
spectra, rather than on actual observations. The first works in 
this field (e.g.\ Buser \& Kurucz 1978; 
Edvardsson \& Bell 1989) were based on sets of synthetic spectra
covering very modest -- by present standards -- intervals of
effective temperature, gravity, and metallicity. The situation 
dramatically improved with the release of a large database of 
ATLAS9 synthetic spectra 
by Kurucz (1993). The first systematic use of Kurucz spectra on
theoretical isochrones has been from Bertelli et al.\ (1994)
and Chiosi et al.\ (1997). 

In order to apply synthetic photometry to sets of theoretical 
isochrones, the basic step consists in the derivation of bolometric 
corrections and temperature--colour relations from the available 
spectra. Several papers deal with the problem (e.g.\ 
Bertelli et al.\ 1994; Chiosi et al.\ 1997; Lejeune et al.\ 1997; 
Bessell et al.\ 1998), 
presenting mathematical formalisms that, although looking somewhat 
different, should be equivalent and produce the same results
when applied to the same sets of spectra, filters, and zero-points.
In the following, we re-write this formalism in a very simple way. 
Our aim is to have generic formulas that, by just 
minimally changing their input quantities, can be applied 
to a wide variety of photometric systems.
%

\subsection{Basic concepts}
\label{sec_outline}

For a star, the spectral flux
as it arrives at the Earth, $f_\lambda$, is simply related 
to the flux at the stellar surface, $F_\lambda$, by
	\beq
f_\lambda = 10^{-0.4 A_\lambda} (R/d)^2 \, F_\lambda \,\,\, ,
	\label{eq_dillution}	
	\eeq
where $R$ is the stellar radius, $d$ is its distance, and $A_\lambda$
is the extinction in magnitudes at the wavelength $\lambda$. 
Once $f_\lambda$ is known, the apparent magnitude $m_{S_\lambda}$,
in a given pass-band with transmission curve $S_\lambda$ comprised
in the interval $[\lambda_1,\lambda_2]$, is given by
	\beq
m_{S_\lambda} = -2.5\,\log\left(
	\frac { \int_{\lambda_1}^{\lambda_2} 
		\lambda f_\lambda S_\lambda \diff\lambda }
		{ \int_{\lambda_1}^{\lambda_2} 
		\lambda f^0_\lambda S_\lambda \diff\lambda } 
	\right)
		+ m^0_{S_\lambda}
	\label{eq_photon}
	\eeq
where $f^0_\lambda$ represents a reference spectrum (not necessarily
a stellar one) that produces a known apparent magnitude $m^0_{S_\lambda}$. 
In other words, $f^0_\lambda$ and $m^0_{S_\lambda}$ completely define 
the zero-points of a synthetic photometric system (see 
Sect.~\ref{sec_zeropoint} below). 

In Eq.~(\ref{eq_photon}), the integrands $\lambda f_\lambda S_\lambda$ 
are proportional to the photon flux 
(i.e.\ number of photons by unit time, surface,
and wavelength interval) at the telescope detector. 
This kind of integration applies well to 
the case of modern photometric systems that have been
defined and calibrated using photon-counting devices such as
CCDs and IR arrays. However, more traditional systems 
like the Johnson-Cousins-Glass $UBVRIJHKLMN$ one, have been
defined using energy-amplifier devices. In this latter case, 
energy integration, i.e.
	\beq
m_{S_\lambda} = -2.5\,\log\left(
	\frac { \int_{\lambda_1}^{\lambda_2} 
		f_\lambda S_\lambda \diff\lambda }
		{ \int_{\lambda_1}^{\lambda_2}
		f^0_\lambda S_\lambda \diff\lambda } 
	\right)
		+ m^0_{S_\lambda} \,\,\, ,
	\label{eq_energy}
	\eeq
would be more appropriate to recover the original system. 
Notice, however, that the difference between energy and photon 
integration is usually very small, unless the pass-bands are 
extremely wide. Unless otherwise stated, in papers of this
series we will adopt the integration of photon counts.

\subsection{Deriving bolometric corrections}
\label{sec_bc}

The starting point of our work are extended libraries of stellar 
intrinsic spectra $F_\lambda$, as derived from atmosphere calculations 
for a grid of effective temperatures \Teff, surface gravities $g$, 
and metallicities $\mh$. The particular library we adopt will be 
described in Sect.~\ref{sec_spectra}. 

From this library, we aim to derive the absolute magnitudes 
$M_{S_\lambda}$ for each star of known $(\Teff, g, \mh)$ -- 
and hence known $F_\lambda$. This can be obtained by means of 
Eq.~(\ref{eq_photon}) (or \ref{eq_energy}), once a distance 
of $d=10$~pc is assumed, i.e.\
	\beq
M_{S_\lambda} = -2.5\,\log\left[
	\left(\!\frac{R}{10\,{\rm pc}}\!\right)^{\!2}
	\!\frac { \int_{\lambda_1}^{\lambda_2} 
		\lambda F_\lambda 10^{-0.4 A_\lambda} 
		S_\lambda \diff\lambda }
		{ \int_{\lambda_1}^{\lambda_2} 
		\lambda f^0_\lambda S_\lambda \diff\lambda } 
	\right]
		+ m^0_{S_\lambda}
	\label{eq_absmag}
	\eeq
and once the stellar radius $R$ is known. 
Since the quantities $(\Teff, g, \mh)$ are not enough to 
specify $R$ (in this case we need to have also the stellar mass $M$), 
we should first eliminate $R$ from our equations. This
is possible if we deal with the bolometric corrections,
	\beqa
BC_{S_\lambda} = M_{\rm bol} - M_{S_\lambda} \,\,\, .
	\label{eq_bc}
	\eeqa
From the definition of bolometric magnitude, 
we have (see also Bessell et al.\ 1998 for a similar approach):
	\beqa
M_{\rm bol} & = & M_{\rm bol, \odot} - 2.5\, \log(L/L_\odot)
	\label{eq_mbol}
	 \\ \nonumber
	 & = & M_{\rm bol, \odot} - 2.5\, \log(
		4\pi R^2 F_{\rm bol} /L_\odot) \,\,\,\, ,
	\eeqa
where $F_{\rm bol} = \int_0^\infty F_\lambda \diff\lambda = 
\sigma T_{\rm eff}^4 $ is the total emerging flux at the stellar
surface.

Substituting Eqs.~(\ref{eq_energy}) and (\ref{eq_mbol}) 
into Eq.~(\ref{eq_bc}), we get
	\beqa
BC_{S_\lambda} & = & M_{\rm bol, \odot} 
	- 2.5\,\log \left[ 
		4\pi (10\,{\rm pc})^2 F_{\rm bol}/L_\odot
		\right] \label{eq_bcfinal} 
	\\ \nonumber
	&& + 2.5\,\log\left(
	\frac { \int_{\lambda_1}^{\lambda_2} 
		\lambda F_\lambda \, 10^{-0.4A_\lambda} 
		S_\lambda \diff\lambda }
		{ \int_{\lambda_1}^{\lambda_2} 
		\lambda f^0_\lambda S_\lambda \diff\lambda } 
	\right)
	- m_{S_\lambda}^0
	\eeqa
that, as expected, depends only the spectral shape
($F_\lambda\,10^{-0.4A_\lambda}/F_{\rm bol}$), and on basic 
astrophysical constants.

To keep consistency with our previous works (e.g.\ Salasnich et al.\
2000), we adopt $M_{\rm bol, \odot}=4.77$, and
$L_\odot=3.844\times10^{33}\,{\rm erg\,s^{-1}}$ (Bahcall et al.\
1995).

By means of Eq.~(\ref{eq_bcfinal}), we tabulate $BC_{S_\lambda}$ for all
spectra in our input library, and for several different photometric
systems. The $BC_{S_\lambda}$ can be then derived for any 
intermediate $(\Teff, g, \mh)$ value, by interpolation 
in the existing grid.  We adopt simple
linear interpolations, with $\log\Teff$, $\log g$, and \mh\ as the
independent variables.  

Next, to attribute absolute magnitudes to stars of given
$(\Teff, L)$ along an isochrone, we simply compute
$M_{\rm bol}$ with Eq.~(\ref{eq_mbol}), and hence
	\beq
M_{S_\lambda} = M_{\rm bol} - BC_{S_\lambda} \,\,\, .
\label{eq_absolmag}
	\eeq

In this formalism, an extinction curve $A_\lambda$ can be applied 
to all spectra of the stellar library, so as to allow the 
derivation of bolometric corrections (and synthetic absolute 
magnitudes) that already include extinction in a 
self-consistent way (cf.\ Eqs.~\ref{eq_bcfinal}
and \ref{eq_absolmag}). With this approach,
the extinction on each pass-band depends not only on the
total amount of extinction, but also on the spectral energy 
distribution of each star -- i.e.\ on its spectral type, luminosity 
class, and metallicity (see also Grebel \& Roberts 1995).
Anyway, for the sake of simplicity, in the present work we will
deal with the case $A_\lambda=0$ only. Tables for different 
extinction curves will be discussed in a forthcoming paper.

Finally, it is worth mentioning that the newly-defined SDSS photometric 
system makes use of an unusual definition for magnitudes (see Lupton 
et al.\ 1999). This specific case, for which some of the above equations do 
not apply, will also be discussed in a subsequent paper of this series. 

\subsection{Reference spectra and zero-points}
\label{sec_zeropoint}

By photometric zero-points, one usually means the constant quantities
that one should add to instrumental magnitudes in order to
transform them to standard magnitudes, for each filter $S_\lambda$.
In the formalism here adopted, however, we do not 
make use of the concept of instrumental magnitude, and hence
such constants do not need to be defined. Throughout this work, 
instead, by ``zero-points'' we refer to the quantities in 
Eqs.~(\ref{eq_photon}) and (\ref{eq_energy}) that depend only 
on the choice of $f^0_\lambda$ and $m^0_{S_\lambda}$. 
They are constant for each filter, and are responsible for
the conversion of the synthetic magnitude scale into a standard 
system. 

As for these quantities, there are four different cases of 
interest:

\subsubsection{VEGAmag systems} 
They make use of Vega ($\alpha$\,Lyr) as
the primary calibrating star. The most famous among these systems is
the Johnson-Cousins-Glass $UBVRIJHKLMN$ one, that can be accurately
recovered by simply assuming that Vega has $V=0.03$~mag, and all 
colours equal to $0$. 
Other systems, like Washington and the HST/WFPC2 VEGAmag one,  
follow a similar definition (some colours, however, are defined to 
have values slightly different from $0$).

Calibrated empirical spectra of Vega are available (e.g.\ Hayes \& Latham 
1975; Hayes 1985), covering the wavelength range from 3300 to
10500~\AA, an interval that can be extended up to 1150~\AA\ when
complemented with IUE spectra (Bohlin et al.\ 1990). 
They can be used to define VEGAmag systems in the optical 
and ultraviolet. However, as the wavelength range accessible
to present instrumentation is much wider, a Vega spectrum covering
the complete spectral range has become necessary. Synthetic spectra
as those computed by Kurucz (1993) and Castelli \& Kurucz (1994),
fulfill this aim. In this case, the predicted fluxes at Vega's 
surface, $F_\lambda$, are scaled by the geometric dilution factor 
	\beq
(R/d)^2=(0.5\,\theta_{\rm d}/206264.81)^2 \,\,\, ,
	\label{eq_diameter}
	\eeq 
where $\theta_{\rm d}$ is the observed Vega's angular diameter 
(in arcsec) corrected by limb darkening. 

More recently, composite spectra of Vega have been constructed by 
assembling empirical and synthetic spectra together 
(e.g.\ Colina et al.\ 1996), so that some small deficiencies 
characteristic of synthetic spectra\footnote{For a discussion 
of the differences between synthetic and observed Vega spectra, 
the reader is referred to Castelli \& Kurucz (1994) and 
Colina et al.\ (1996).} are corrected. 
This has the precise scope of providing a reference spectrum for
conversions between apparent magnitudes of real (observed)
stars, and physical fluxes. 
However, it is not clear whether such composite Vega 
spectrum should be preferable when synthetic photometry
is performed on theoretical spectra, as in the present work.
For instance, if the ATLAS9 spectrum for Vega has the core of
Balmer lines differing by as much as
10 percent from the observed ones (cf.\ Colina et al.\ 1996),
it is probable that the same deviations will be present in all Kurucz
spectra of comparable temperatures/gravities. If this is the case, 
the synthetic Vega spectrum would probably give better zero-points 
for these stars than the composite Colina et al.\ (1996) one.

For this reason, we simply adopt the synthetic ATLAS9 
model for Vega, with $\Teff=9550$~K, $\log g=3.95$, 
$[{\rm M/H}]=-0.5$, 
and microturbulent velocity $\xi=2\,{\rm km\,s}^{-1}$, 
the spectrum being provided by Kurucz (1993). 
Castelli \& Kurucz (1994) have computed a higher-resolution spectrum
for the same model, using the more refined ATLAS12 code. As 
discussed by these authors, ATLAS9 and ATLAS12 spectra for 
Vega are almost identical.

Once the synthetic model $F_\lambda^{\rm Vega}$ is chosen, 
we just need to adopt a fixed value for the dilution factor 
$(R/d)^2$, in order to have $f_\lambda^{\rm Vega}$ at the 
Earth's surface (outside the atmosphere). 
Two choices are possible then: We
either (i) adopt an observed value of 
$\theta_{\rm d}$ as input to Eq.~(\ref{eq_diameter}),
or (ii) adopt the observed Vega flux as measured at the
stellar surface, at a given wavelength, as an input to 
Eq.~(\ref{eq_dillution}).

Since direct measures of $\theta_{\rm d}$ are relatively more
uncertain than direct measures of $f_\lambda^{\rm Vega}$,
we prefer to adopt the second alternative: Taking the flux values
at 5556~\AA\ from Hayes (1985; 
$f_\nu^{\rm Vega}=3.542\times10^{-20}$ ${\rm erg \, 
s^{-1} \, cm^{-2} \, Hz^{-1}}$), and at 5550~\AA\ from 
Kurucz (1993) Vega model 
($F_\lambda^{\rm Vega}=5.507\times10^7$ ${\rm erg \, 
s^{-1} \, cm^{-2} \, \AA^{-1}}$), 
we obtain $(R/d)^2=6.247\times10^{-17}$. This 
value implies an angular diameter of $\theta_{\rm d}=3.26$~mas
(Eq.~\ref{eq_diameter}) for Vega,
that compares very well with the observed values of
$3.24\pm0.07$~mas (Code et al.\ 1976) and
$3.28\pm0.06$~mas (Ciardi et al.\ 2000).

It is worth remarking that, since the zero-points in
VEGAmag systems are attached
to the observed Vega fluxes, their synthetic absolute magnitudes 
may have systematic errors of a few hundredths of
magnitude (say up to 0.03~mag), which is the typical magnitude of 
errors in measuring fluxes at the Earth's surface. Somewhat smaller 
errors, however, are expected in the colours. 
These uncertainties will probably not be eliminated unless 
definitive $\theta_{\rm d}$ and $f_\lambda^{\rm Vega}$ 
measurements become available.

\subsubsection{ABmag systems} 
In the original work by Oke (1964), monochromatic AB magnitudes are 
defined by
	\beq
m_{{\rm AB}, \nu} = -2.5\,\log f_\nu - 48.60 \,\,\,.
	\label{eq_abmagdef}
	\eeq
This means that a reference spectrum of 
constant flux density per unit frequency
	\beq
f_{{\rm AB},\nu}^0=3.631\times10^{-20}\,\,\, {\rm erg \, 
s^{-1} \, cm^{-2} \, Hz^{-1}} 
	\label{eq_abmagzp}
	\eeq
will have AB magnitudes $m_{{\rm AB}, \nu}^0=0$ at all frequencies $\nu$. 

This definition can be extended to any filter system, provided that 
we replace the monochromatic flux $f_\nu$ with the 
photon counts over each pass-band $S_\lambda$ obtained from the star, 
compared to the photon counts that one would get by observing 
$f_{{\rm AB},\nu}^0$:
	\beq
m_{{\rm AB}, S_\lambda} = -2.5\,\log 
	\left[
		\frac{ \int_{\lambda_1}^{\lambda_2} 
		(\lambda/hc) f_\lambda S_\lambda\diff\lambda }
		{ \int_{\lambda_1}^{\lambda_2} 
		(\lambda/hc) f_{{\rm AB},\lambda}^0 
			S_\lambda\diff\lambda }
	\right] \,\,\,,
	\label{eq_abmag}
	\eeq
where $f_{{\rm AB},\lambda}^0 = f_{{\rm AB},\nu}^0\,c/\lambda^2$. 
Then, it is easy to show that 
Eq.~(\ref{eq_abmag}) is is just a particular case of 
our former Eq.~(\ref{eq_photon}), for which $f_{{\rm AB},\nu}^0$
is the reference spectrum, and $m_{{\rm AB}, S_\lambda}^0=0$ are the 
reference magnitudes.

It is worth mentioning that different practical implementations of
the ABmag system have been defined over the years (e.g.\ Oke \& Gunn
1979, and Fukugita et al.\ 1996). 
They differ only in the definition of the reference 
stars (or reference stellar spectra) used as spectrophotometric 
standards during the conversion from {\em observed} 
instrumental magnitudes into fluxes $f_\nu$ (that are used 
in Eq.~\ref{eq_abmagdef}). Since we are dealing
with synthetic spectra only, such a conversion is not necessary in our 
case. It follows that these different definitions are not a
point of concern to us.

\subsubsection{STmag systems} 
ST monochromatic magnitudes have been introduced
by the HST team, and are defined by
	\beq
m_{\rm ST, \lambda} = -2.5\,\log f_\lambda - 21.10 \,\,\,.
	\label{eq_stmagdef}
	\eeq
This means that a reference spectrum of 
constant flux density per unit wavelength
	\beq 
f_{\rm ST,\lambda}^0=3.631\times10^{-9} \,\,\, {\rm erg \, 
s^{-1} \, cm^{-2} \, \AA^{-1}} 
	\label{eq_stmagzp}
	\eeq
will have ST magnitudes $m_{\rm ST, \lambda}^0=0$ at all wavelengths. 
Similarly to the case of AB magnitudes, this can be generalized to any 
pass-band system with:
	\beq
m_{{\rm ST}, S_\lambda} = -2.5\,\log 
	\left[
		\frac{ \int_{\lambda_1}^{\lambda_2} 
		(\lambda/hc) f_\lambda S_\lambda\diff\lambda }
		{ \int_{\lambda_1}^{\lambda_2} 
		(\lambda/hc) f_{\rm ST,\lambda}^0 S_\lambda\diff\lambda }
	\right] \,\,\,.
	\label{eq_stmag}
	\eeq

Again, this situation is easily reproduced in our formalism 
with the adoption of $f_\lambda^0 = f_{\rm ST,\lambda}^0$ 
and $m_{{\rm ST}, S_\lambda}^0=0$.

\subsubsection{``Standard stars'' systems} 
This class comprises 
all photometric systems that use standard stars different from
Vega to define the zero-points. Good examples are the Vilnius 
(Straizys \& Zdanavicius 1965) and Thuan-Gunn (Thuan \& Gunn 1976) 
systems. 

In this case, we are forced to use empirical spectra of standard
stars in order to define $f^0_\lambda$ and $m^0_{S_\lambda}$. 
A good set is provided by the four metal-poor subdwarfs 
BD\,+17\degr4708, BD+26\degr2606, HD\,19445 and HD\,84937, which 
are widely-used spectrophometric secondary standards 
(Oke \& Gunn 1983), as well as standards for several photometric 
systems. 

Actually, the present work does not deal with any ``standard stars'' 
system, and this case is here included just for the sake of
completeness. 
Details about a few specific systems -- including the possible 
choices for $f^0_\lambda$ -- will be given in forthcoming papers.

\section{The stellar spectral library}
\label{sec_spectra}

\begin{figure}
	\resizebox{0.9\hsize}{!}{\includegraphics{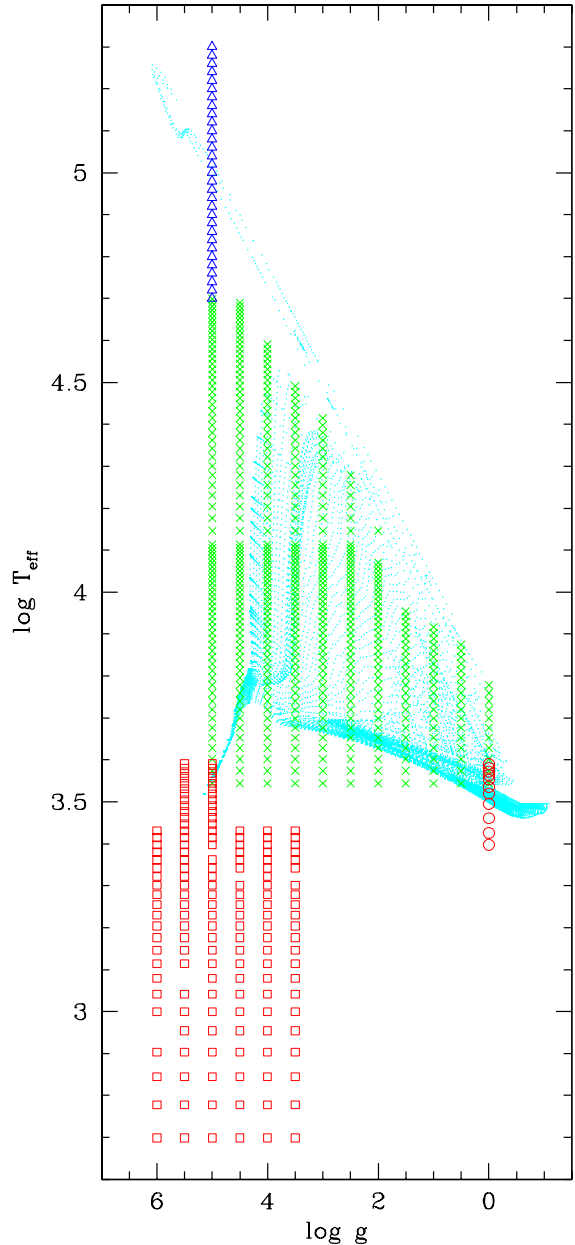}}
	\caption{Distribution of the  $\mh=0$ spectra incorporated in our
stellar library (large symbols) in the $\log\Teff - \log g$ plane, 
compared to the position of stellar models of solar metallicity
(small dots; these models are isochrones to be discussed later in
Sect.~\ref{sec_basic}). The spectra are taken from Castelli et al.\ 
(1997; crosses), Fluks et al.\ (1994; circles), 
Allard et al.\ (2000a; squares), and pure blackbody
(triangles). Fluks et al.\ spectra have been arbitrarily located at
$\log g=0$. Similar
distributions hold for all metallicities between $\mh=-2.5$ and $+0.5$
(see text).}
	\label{fig_griglia}
	\end{figure}

In the following, we will present the stellar spectral library
put together for this work. For the sake of reference, 
Fig.~\ref{fig_griglia} presents the distribution of all spectra 
in the $\log\Teff - \log g$ plane.

\subsection{Kurucz atmospheres}
\label{sec_kurucz}

Earlier Padova isochrones were based on the Kurucz (1993) 
libraries of ATLAS9 synthetic atmospheres. 
As discussed in a series
of papers by Castelli et al.\ (1997), Bessell et al.\ (1998),
and Castelli (1999), these models are superseded by now.
Firstly, small discontinuities associated to the scheme of
``approximate overshooting'' initially adopted by Kurucz have 
been corrected (cf.\ Bessell et al.\ 1998). 
Secondly, no-overshooting
models have been demonstrated to produce \Teff-colour relations
in better agreement with empirical ones, at least for stars hotter
than the Sun (Castelli et al.\ 1997).

\subsubsection{The more recent models}

In the present work, we adopt the ATLAS9
no-overshoot models that have been calculated by 
Castelli et al.\ (1997). They correspond to the ``NOVER'' 
files available at \verb$http://cfaku5.harvard.edu/grids.html$. 
The metallicities cover the values $\mh=-2.5$, $-2.0$, $-1.5$, 
$-1.0$, $-0.5$, $0.0$, and $+0.5$, with solar-scaled abundance ratios. 
A microturbulent velocity $\xi=2\,{\rm km\,s^{-1}}$, and a mixing length 
parameter $\alpha=1.25$, are adopted.
Notice that these models are now being extended so as to include
also $\alpha$-enhanced chemical mixtures, which represents a potentially
important improvement for our future works.

Kurucz models cover quite well the region of the 
$\log\Teff$ vs.\ $\log g$ plane actually occupied by stars,
at least in the $3500\,{\rm K} \le T \le 50\,000\,{\rm K}$,
$0\le \log g \le5$ intervals (see Fig.~\ref{fig_griglia}). 
However, it has to be extended to
both lower and higher \Teff s, as will be detailed below.

It is important to recall that Kurucz (ATLAS9) spectra are widely 
used in the field of synthetic photometry, mainly because of their
wide coverage of stellar parameters and easy availability. 
Moreover, there are also good indications in the literature that
these spectra do a good job in synthetic photometry, 
provided that we are dealing with broad-band systems.
Compelling examples of this can be found in 
Bessell et al. (1998), who compares the $UBVRIJHKL$ results obtained 
from the recent ATLAS9 spectra to empirical relations derived with the 
infrared flux method, lunar occultations, interferometry, and eclipsing
binaries. Their results indicate that the 1998 ATLAS9 models are 
well suited to synthetic photometry, but for small errors, 
generally lower than 0.1 mag in colours, that we do not 
consider as critical. In fact, we are more interested in the 
overall dependencies of colours and magnitudes with stellar 
parameters -- probably well represented by present synthetic 
spectra -- than on details of this order of magnitude.

Additionally, Worthey (1994) presented extensive comparisons between 
Kurucz (1993) spectra and stars in the low-resolution spectral 
library by Gunn \& Stryker (1983), obtaining generally a good match
for wavelengths redder than the $B$ pass-band. Worthey's figure 9 
also presents a comparison between Kurucz (1993) solar spectra and 
Neckel \& Labs (1984) data, with excellent results (errors 
lower than 0.1 mag) all the way from the UV up to the near-IR. 
Since the ATLAS9 1998 spectra differ just little from the Kurucz 
(1993) version (a few percent in extreme cases), these
results are to be considered still valid. 

\subsubsection{Some caveats on ATLAS9 spectra}

The previously mentioned works point to a reasonably good 
agreement between ATLAS9 spectra and those of real stars of
near-solar metallicity,
especially in the visual and near-infrared pass-bands. 
However, there are many known inadequacies in these 
spectra, which should be kept in mind as well. 
Here, we give just a brief list of the potential problems, 
concentrating on those which may be more affecting our 
synthetic colours.

ATLAS9 spectra are based on 1D static and plan-parallel 
LTE model atmospheres, which use a huge database of atomic  
line data (Kurucz 1995). The line list is known not to be 
accurate: In fact, Bell et al.\ (1994) show 
that the solar spectra calculated using Kurucz 
list of atomic data present many unobserved lines; 
moreover, the number of lines which are too strong exceeds 
those which are too weak. 
The problem can be appreciated by looking at the high-resolution
spectral plots presented by Bell et al.\ (1994), but could 
hardly be noticeable in low-resolution plots (such as in the
comparisons presented in Worthey's 1994 figure 9, 
and in Castelli et al.\ 1997 figure 2). 

Also, Bell et al.\ (2001) show that a motivated increase in
the Fe\,{\sc i} bound-free opacity cause a significant 
improvement in the fitting of the solar spectrum in the
$3000-4000$~\AA\ wavelength region, affecting the entire 
UV region as well. Such increased sources of continuous 
opacity are still missing in ATLAS9 atmospheres\footnote{In 
the original Kurucz (1993) spectra, a modest 
reduction of the continuous UV flux results from the
adopted ``approximate overshooting'' scheme.}.

These results indicate that 
ATLAS9 spectra will produce worse results when applied to
(i) narrow-band photometric systems, in which individual metallic 
lines can more significantly affect the colours, and (ii) in the UV 
region, especially shortward of 2720 \AA\ (see Bell et al.\ 2001). 
In both cases, the errors caused by wrong atomica data are such that 
we can expect not only systematic and \Teff-dependent
offsets in synthetic colours,
but also a somewhat wrong dependence on metallicity.
Clearly, these points are worth being properly investigated by
means of detailed spectral comparisons.

Regarding the present work, the above-mentioned problems (i) 
critically determine the inadequacy of synthetic colours computed
for the Str\"omgren system (Girardi et al., in preparation), 
and (ii) may possibly cause significant errors in our synthetic 
HST/WFPC2 UV colours.

Other potential problems worth of mention are:
	\begin{itemize}
	\item 
The deficiencies in 1D atmosphere models and in the mixing length 
description of convection, as compared to exploratory 3D model 
atmospheres (e.g.\ Asplund et al.\ 2000). 
They should affect mainly the intermediate-strong lines with 
significant non-thermal broadening and in stars cooler than 
$\sim6000$~K. The consequences of spatial and thermal inhomogeneities 
in the atmosphere of the Sun and Procyon have been investigated by
Allende Prieto et al.\ (2001, 2002). 
	\item
Inadequacies in the hydrogen line computations. Especially 
affected are high members of the Balmer series in GK stars 
-- due to inhomogeneities in real
atmospheres and inadequate treatment of hydrogen line broadening 
in cool stars (see Barklem et al.\ 2000a,b)
-- and the core of Balmer lines and the region around the Paschen 
discontinuity and longward in A stars (Colina et al.\ 1996).
In synthetic photometry, these inadequacies are
expected to slightly affect the $U$ and $B$ magnitudes for
A-F stars, but to have a greater effect in Str\"omgren indices.
	\item 
The inadequacy of using scaled-solar metal ratios
in giants where the abundance of CNO elements has been altered by
dredge-up events. Similarly, the inadequancy of using scaled-solar
abundances for halo and bulge stars, instead of $\alpha$-enhanced 
ones.
	\end{itemize}

Obviously,
some improvement upon these points is expected in future releases
of ATLAS spectra (see Castelli \& Kurucz 2001), and of other
extended spectral grids as well.
Fortunately, the work by Bessell et al.\ (1998) 
gives us some confidence that present broad-band magnitudes 
and colours (from $U$ to $K$) are modelled with an accuracy
that is already acceptable for many applications.

Finally, we remark that some authors (Lejeune et al.\ 1997, 1998) 
propose the application of {\em a posteriori} transformations
to Kurucz (1993) spectra, as a function of wavelength and \Teff,
such as to reduce the errors of the derived synthetic $UBVRIJHKL$ 
photometry. In our opinion, 
such transformations are questionable because they do 
not correct the cause of the discrepancies -- majorly 
identifiable in the imperfect modelling of absorption lines --
and the case for applying them to stars of all surface 
metallicities and gravities is far from compelling.

\subsection{Extension to higher temperatures}
\label{sec_blackbody}

For $\Teff>50\,000$~K, we simply assume black-body spectra. 
This is probably a good approximation for wavelengths 
$\lambda>912$~\AA. In fact, we find always a reasonably smooth
transition in the computed $BC_{S_\lambda}$s as we cross the
$\Teff=50\,000$~K temperature boundary.

\subsection{Extension to M giants}
\label{sec_mgiants}

Synthetic spectra for M giants have still many problems -- 
mainly in their ultraviolet-blue region -- that 
partially derive from incomplete opacity lists of molecules
such as TiO, VO and H$_2$O (see e.g.\
Plez 1999; Alvarez \& Plez 1998; Alvarez et al.\ 2000; 
and Houdashelt et al.\ 2000a,b to appreciate the state of 
the art in the field).

Therefore, we prefer to use the empirical M giant spectra
from Fluks et al.\ (1994; or ``intrinsic'' spectra as referred 
in their paper). They cover the wavelength interval from 3800~\AA\ 
to 9000~\AA. Outside this interval, the empirical spectra have been
extendend with the ``best fit'' synthetic spectra computed by the
same authors.

\begin{figure*}
	\resizebox{\hsize}{!}{\includegraphics{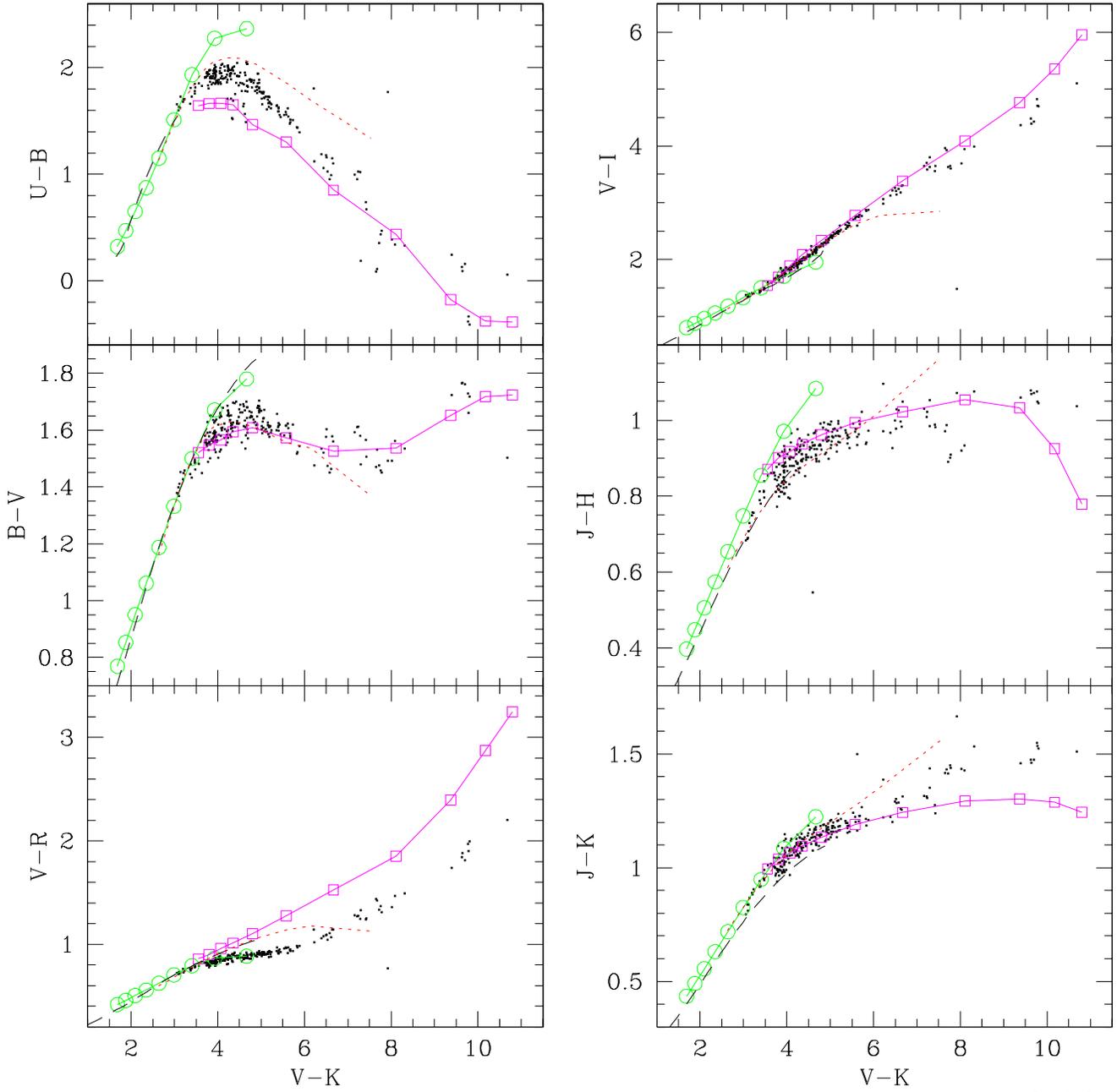}}
	\caption{Colour vs.\ \vk\ relations for giants. 
The connected open circles represent the 
relation obtained from $\mh=0$ ATLAS9 spectra located 
along the $\Teff = 3250 + 500\,\log g$ line (the typical location for
RGB stars) in the diagram of Fig.~\protect\ref{fig_griglia}. 
The connected open squares correspond to the 
relation obtained from the M-giant spectra from Fluks et al.\ (1994), 
completed and modified at $\lambda<4800$~\AA\ as detailed in the text.
The dashed lines represent the empirical relations for F0--K5 
solar-metallicity giants from Alonso et al.\ (1999b), {whereas the
dotted lines are the synthetic relations for K0--M7 giants from
Houdashelt et al.\ 2000a)}. 
The empirical data for M giants (small dots) are from 
Fluks et al.\ (1994). As far as possible, all observations have been 
converted to the same photometric system as used in our synthetic 
photometry (i.e. the ``Bessell'' $UBVRIJHK$ system; see text). }
	\label{fig_colori_bv}
	\end{figure*}

However, the whole procedure reveals a problem: 
if we simply merge empirical and synthetic spectra from 
Fluks et al.\ (1994), the resulting synthetic \bv\ and \ub\ colours 
just badly correlate with the measured colours for the 
same stars (which were also obtained by Fluks et al.\ 1994). 
This problem probably derives from a
bad flux calibration at the blue extremity
of the observed spectra and/or from the imperfect match between
synthetic and observed spectra at 3800~\AA. In order
to circumvent (at least partially) the problem, we simply
multiply each M-giant spectrum blueward of 4000~\AA\ (with a smooth 
transition in the range from 4000~\AA\ to 4800~\AA) by a constant,
typically between 0.8 and 1.2, so that the synthetic colours recover 
the observed behaviour of the \bv\ vs.\ \vk\ data. 
The first two panels of Fig.~\ref{fig_colori_bv} show the results. 

Actually, Fig.~\ref{fig_colori_bv} presents six different 
colour vs.\ \vk\ diagrams that are useful to understand the 
situation for giants. Care has been taken in expressing data and models 
in the same photometric system, the ``Bessell''
$UBVRIJHK$ one, that we will detail later in Sect.~\ref{sec_johnson}.
For M giants, the empirical photometric data from Fluks et al.\ 
(1994; small dots) can be compared with the results of our 
synthetic photometry\footnote{The $JHK$ colours 
from Fluks et al.\ are in the ESO system and have 
been converted to the Bessell one by using the relations 
found in Bessell \& Brett (1988).}. Noteworthy, there is a 
reasonably good match between the synthetic and observed relations
for most colours. This has been imposed for \ub\ and \bv, whereas is
a natural result for all colours involving wavelengths longer 
than $\sim4800$~\AA. The only clear exception is the \vr\ colour, 
for which differences of $\sim0.4$~mag are found for all giants 
of spectral type later than M4 ($\vk\ga5$). The reason for this 
discrepancy is not clear, but may lie in the use of $R$ filters 
with different transmission curves. Also the predictions for
\jk\ do not fit well all the photometric data, somewhat failing
for the spectral types later than M7 ($\vk\ga8$). However, since 
these latters are quite rare, such mismatch does not pose a 
serious problem.

For the sake of comparison, Fig.~2 
also presents the relations obtained by means of
the M-giant models from Houdashelt et al.~(2000a), in the
case of solar metallicity. Together with other recent examples
(e.g.\ Plez 1999; Alvarez et al.\ 2000), they represent 
state-of-the-art computations of cool oxygen-rich stellar 
atmospheres. As can be appreciated in the figure, 
Houdashelt et al.\ models reproduce well the empirical data as 
far as $V\!-\!K\la6$ (spectral types earlier than M5), but start 
departing from these for cooler stars.  A similar situation 
holds if we look at different $\Teff$--colour relations, as
can be seen in figures 13 and 14 of Houdashelt et al. (2000a),
where they compare their $\Teff$--colour relations with those
obtained with Fluks et al.\ (1994) spectra and data for field 
giants. Also in this case, it seems that Fluks et al.\ (1994)
spectra do better reproduce the empirical relations for the 
spectral types later than M4. 

Once we have defined the library of M-giant spectra, 
we associate effective temperatures to them by using the scale 
favoured by Fluks et al.\ (1994). In this scale, M giants cover 
the temperature interval from $3\,850$~K 
(MK type M0) to $2\,500$~K (MK type M10).
We recall that Fluks et al.\ (1994) \Teff\ values
are derived from a careful fitting of the observed spectra with 
synthetic model atmospheres of solar metallicity. Their scale is also 
in excellent agreement with the empirical one from Ridgway et al.\ 
(1980), which covers spectral types earlier than M6. 

After the proper \Teff\ is attributed, each one of our modified 
spectra is completely re-scaled by a constant, so that the 
total flux vs.\ \Teff\ relation -- i.e.\ 
$F_{\rm bol}=\sigma T_{\rm eff}^4$ -- is recovered.

Finally, we face the problem of defining the transition between 
the M-giant spectra, and the ATLAS9 ones which are available for
temperatures higher than $3\,500$~K. To this aim, it is helpful
to examine Fig.~\ref{fig_colori_bv}, where we also include: 
	\begin{itemize} 
	\item
the synthetic photometry for a sequence of ATLAS9 spectra for 
M giants (empty circles). These
spectra are located along the line $\Teff = 3250 + 500\,\log g$,
which represents fairly well the locus of low-mass red giants 
in a $\Teff$ vs.\ $\log g$ diagram (see Fig.~\ref{fig_griglia}).
	\item
the mean empirical relations for F0--K5 giants of $\feh=0$
as derived from Alonso et al.\ (1999b) fitting formulas 
(dashed lines)\footnote{Alonso et al.\ (1999b) \vi\ and \vr\ 
colours have been transformed from Johnson to Johnson-Cousins systems 
using relations from Bessell (1979), whereas infrared colours
were converted from TCS to ``Bessell'' systems using the relations
from Alonso et al.\ (1998).}.   
	\end{itemize} 
An important fact to be noticed is that our synthetic photometry
reproduces Alonso et al.\ (1999b) relations
in all colours remarkably well. 

From inspecting this and other similar plots, we can conclude that
the mismatch between Kurucz ATLAS9 and 
Fluks et al.\ (1994) spectra starts at about $\Teff=3\,850$ and
increases slowly as the temperature decreases down to $3\,500$~K 
(i.e.\ from $\vk\simeq3.5$ to $\vk\simeq4.7$). 
Hence, we adopt a smooth transition between these two spectral 
sources over this temperature interval. The same
M giant spectra are assumed for all metallicities.

The complete procedure ensures reasonable colour vs.\ \vk\ relations
for all giants of near-solar 
metallicity (Fig.~\ref{fig_colori_bv}). Nevertheless, this kind
of approach cannot be completely satisfactory, first because the 
original Fluks et al.\ (1994) spectra have been artificially 
corrected at wavelengths shorter than 4800~\AA\ in order to produce
reasonable \bv\ and \ub, 
and second because we do not dispose of similar M-giant spectra for 
metallicities very different from solar. Better empirical and 
theoretical spectra for M giants seem to be urgently needed.
Anyway, in the context of the present work the problem is not 
dramatic because M giants cooler than $\Teff\sim3\,500$~K are only 
found in the RGB-tip and TP-AGB phases of high metallicity stellar 
populations, and constitute just a tiny fraction of the number of
red giants.  The problem could be critical, instead, when we consider 
integrated properties of stellar populations, because M giants,
despite their small numbers, have high luminosities and
contribute a sizeable fraction of the integrated light. 

\subsection{Extension to M+L+T dwarfs}
\label{sec_mdwarfs}

Although the modelling of cool dwarfs atmospheres presents 
challenges comparable to those found in late-M giants 
(e.g.\ the inadequacy of TiO and H$_2$O line lists, and dust 
formation; see Tsuji et al.\ 1996, 1999; Leggett et al.\ 2000), 
present results compare reasonably well 
with observational spectral data (see e.g.\ figure 9 in
both Leggett et al.\ 2000 and 2001). A review on the subject
can be found in Allard et al.\ (1997).

An extended library of synthetic spectra for cool dwarfs (of types M 
and later) is provided by Allard et al.\ (2000a; see
\verb$ftp://ftp.ens-Lyon.fr/pub/users/CRAL/fallard$). 
We use their set of 
``BDdusty1999'' atmospheres (see also Chabrier et al.\ 
2000; Allard et al.\ 2000b, 2001), 
that should supersede the ``NextGen'' models from the same
group (Hauschildt et al.\ 1999) due to the consideration of 
better opacity lists and dust 
formation. Dust can significantly affect the coolest atmospheres,
corresponding to dwarfs of spectral types L and T.

The selected spectra cover the \Teff\ intervals:
	\begin{itemize}
	\item from $4\,000$~K to $2\,800$~K (``AMES'' models)
for metallicities $\mh=0.0$, $-0.5$, $-1.0$, $-1.5$, and both 
$\log g=5.0$ and $5.5$;
	\item from $2\,800$~K down to $500$~K (``AMES-dusty'' 
models) only for $\mh=0.0$, and $\log g$ values between $3.5$ and $6.0$.
	\end{itemize}
These spectra are presented with a extremely high resolution, that by
far exceeds the one necessary in our work. Thus, we have convolved the 
flux per unit frequency $F_\nu$ with a Gaussian filter of
$\sigma_\nu=2.4\times10^{-18}\, {\rm Hz}$, that corresponds to a FWHM
of 20~\AA\ at $\lambda=5550$~\AA.  The resulting spectra were then
reported to the same grid of wavelengths of Kurucz' spectra. 

We find that there is a good agreement between 
ATLAS9 and BDdusty1999 spectra in the \Teff\ range between 
$\sim3\,800$~K and $4\,000$~K. Then, we set the transition between 
ATLAS9 and BDdusty1999 spectra at $\sim3\,900$~K. This choice 
guarantees smooth \Teff\ vs.\ colour relations for dwarfs.

\section{Photometric systems available}
\label{sec_systems}

\begin{figure*}
	\resizebox{\hsize}{!}{\includegraphics{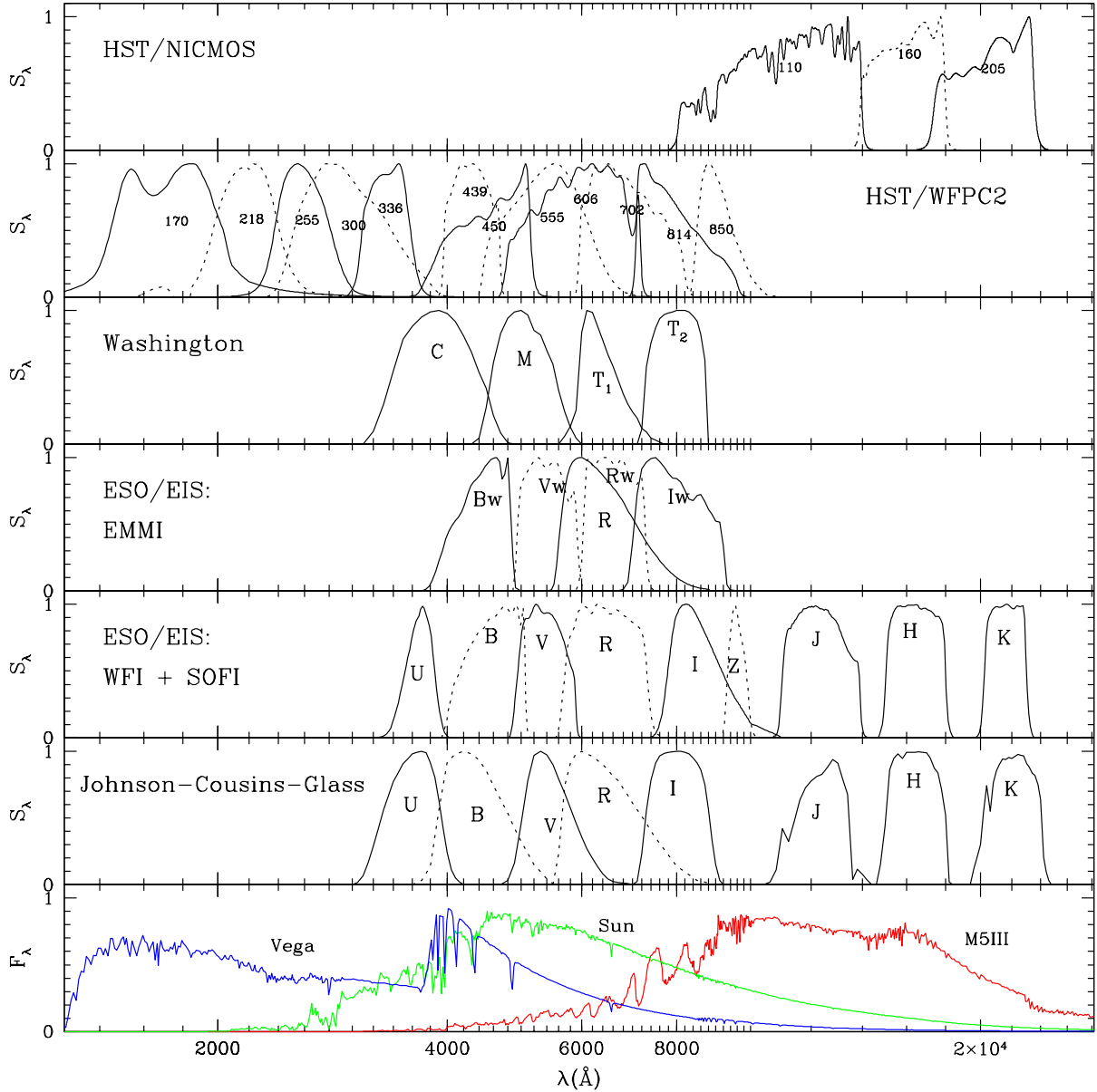}}
	\caption{The filter sets used in the
present work. From top to bottom, we show the
filter+detector transmission curves $S_\lambda$ for the systems: 
(1) HST/NICMOS, (2) HST/WFPC2, (3) Washington, (4) ESO/EMMI,
(5) ESO/WFI $UBVRIZ$ + ESO/SOFI $JHK$, and (6) Johnson-Cousins-Glass. 
All references are given in Sect.~\protect\ref{sec_systems}. 
To allow a good visualisation of the filter curves, 
they have been re-normalized to their 
maximum value of $S_\lambda$.
For the sake of comparison, the bottom panel presents 
the spectra of Vega (A0V), the Sun (G2V), and a M5 giant,
in arbitrary scales of $F_\lambda$.}
	\label{fig_filtri}
	\end{figure*}

In this section we present the basic information regarding the filter
transmission curves and zero-points for each photometric system.
As a reference to the discussion, Fig.~\ref{fig_filtri} presents 
the filter sets under consideration, as compared to the spectra of a
hot (Vega), an intermediate (the Sun), and a cool star (an M5 giant).

\subsection{Johnson-Cousins-Glass system}
\label{sec_johnson}

Aiming to reproduce the Johnson-Cousins-Glass system,
we adopt the filter pass-bands indicated by Bessell (1990; for 
Johnson-Cousins $UBVRI$) and Bessell \& Brett (1988; for $JHK$). 
We also apply their prescription for computing the $U-B$ colour 
by means of a slightly modified pass-band $BX_{90}$, instead of 
the normal $B$ one. Moreover, in order to better recover the 
original system, we adopt energy instead of photon count
integrations (see Sect.~\ref{sec_outline}). 

It is worth recalling that these pass-bands  
represent just one specific version
of the ``standard'' Johnson-Cousins-Glass system, that may 
differ from filter systems in usage at several observatories.
The Bessell \& Brett (1988) $JHKLM$ pass-bands, for instance, 
represent an effort in the direction of homogenizing 
several different near-infrared systems 
(SAAO, ESO, CIT/CTIO, MSO, AAO, and Arizona). 
In Bessell (1990) and 
Bessell \& Brett (1988), the reader can find a set of
useful fitting relations between colours in the several 
original systems.

As previously mentioned, Johnson-Cousins-Glass is essentially a 
VEGAmag system. We fix the zero-points by assuming that Vega has 
apparent magnitudes equal to $0.03$ in all $UBVRIJHK$ bands, i.e.\ 
we impose all colours to be null. Notice that our definition is
very similar to the Bessell et al.\ (1998) one, who adopt Vega 
colours differing from zero by just some thousandths of a magnitude
(see their table A1).

\subsection{Instruments on board of HST}

In the context of the present work, the distinctive feature 
of HST photometry is the use of ABmag and STmag systems, which 
greatly simplifies the definition of zero-points. WFPC2 and NICMOS
observations can also be expressed in VEGAmag magnitudes.

\subsubsection{WFPC2}
\label{sec_wfpc2}

As for the WFPC2, we produce bolometric corrections and magnitudes 
in the F170W, F218W, F255W, F300W, F336W, F439W, F450W, F555W, 
F606W, F702W, F814W, and F850LP filters. Similar tables have been
produced in previous papers (Chiosi et al.\ 1997; 
Salasnich et al.\ 2000); the
present ones differ just in minor details.

The transformations have been computed in STmag, ABmag, and VEGAmag
systems. Whereas the STmag and ABmag systems can be
straightforwardly simulated, the VEGAmag one deserves some
comments.

According to the SYNPHOT package distributed with the STSDAS 
software, in a VEGAmag system Vega should have apparent magnitudes 
equal to zero in all pass-bands. However,
the most widely used calibration of WFPC2 photometry
comes from Holtzman et al.\ (1995), who adopt a set of zero-points
in which Vega has apparent magnitudes
$U=0.02$, $B=0.02$, $V=0.03$, $R=0.039$, $I=0.035$. 
Accordingly, we choose this latter definition, and 
impose that for the WFPC2 filters that correspond to $UBVRI$ 
in wavelength, the synthetic magnitudes of Vega have these 
same values. For filters of intermediate wavelength, 
we adopt a linear interpolation between these values, 
whereas for bluer and redder filters a Vega magnitude equal 
to zero is assumed. 

As for the $S_\lambda$ functions, we use the
pre-launch pass-bands kindly provided by Jon Holtzman 
(see Holtzman et al.\ 1995). We recall that,
owing to the presence of contaminants inside the WFPC2 
(see Baggett \& Gonzaga  1998; Holtzman et al.\ 1995), $S_\lambda$
changes slowly with time, especially for UV filters. 
To cope with this, observers usually apply small corrections 
to the definition of the instrumental magnitudes,  
in order to bring the magnitudes back to the original conditions
(see Holtzman et al.\ 1995). This justifies our use of
pre-launch pass-bands instead of the present-day ones
provided by SYNPHOT.

\subsubsection{NICMOS}
\label{sec_nicmos}

For NICMOS filters, we compute bolometric corrections and 
absolute magnitudes in the ABmag, STmag, and VEGAmag systems. 
For the moment, calculations are limited to the
three most frequently used filters, i.e.\ F110W, F160W, and F205W.
The pass-bands come from SYNPHOT, and have been kindly provided
by Don Figer.

NICMOS observations are also frequently expressed in units
of milli-Jansky (mJy). The conversion between AB magnitudes and 
mJy is straightforward:
	\beq
m_{{\rm AB}, \nu} ({\rm mag}) = -2.5 \log
	f_\nu({\rm mJy}) - 16.4 
	\,\,.
	\eeq

\subsection{ESO Imaging Survey (EIS) filter sets} 
\label{sec_eis}

The EIS survey (Renzini \& da Costa 1997; da Costa 2000) 
aims at providing a large database of 
deep photometric data, among which the astronomical community could 
select interesting targets for VLT spectroscopy. 
The survey is conducted at several different instruments at ESO:
 
\subsubsection{WFI}
\label{sec_wfi}

The Wide Field Imager (WFI) at the 
MPG/ESO 2.2m La Silla telescope provides imaging of excellent 
quality over a $34^\prime\times33^\prime$ field of view.
It contains a peculiar set of broad-band filters, 
very different from the ``standard'' Johnson-Cousins ones.
This can be appreciated in Fig.~\ref{fig_filtri}; notice in
particular the particular shapes of the WFI $B$ and $I$ 
filters. Moreover, EIS makes use of the WFI $Z$ filter which 
does not have a correspondency in the Johnson-Cousins system.

Given the very unusual set of filters, the importance of 
computing isochrones specific for WFI is evident. This has benn 
done so for the broad WFI filters $U$ (ESO\#841), $B$ (ESO\#842),
$V$ (ESO\#843), $R$ (ESO\#844), $I$ (ESO\#845), and $Z$ (ESO\#846), 
that -- here and in Fig.~\ref{fig_filtri} -- are referred to 
as $UBVRIZ$ for short.

Bolometric corrections have been computed in the VEGAmag 
system assuming all Vega apparent magnitudes to be $0.03$,
and in the ABmag system, which is adopted by the EIS group. 
The photometric calibration
of EIS data is discussed in Arnouts et al.\ (2001). 

It is very important to notice that any 
photometric observation performed with WFI that makes use
of standard stars (e.g.\ Landolt 1992) to convert WFI instrumental
magnitudes to the standard Johnson-Cousins $UBVRI$ system,
{\em will not be in the WFI VEGAmag system we are dealing
with here}. Instead, in that case, we should better
compare the observations with the isochrones in the
standard Johnson-Cousins system, or, alternatively, apply
colour transformations to our WFI isochrones so that they
reproduce the \Teff-colour sequences of dwarfs and giants
in the Johnson-Cousins system. This aspect will be better 
illustrated Sect.~\ref{sec_conclu}. 

\subsubsection{SOFI}
\label{sec_sofi}

SOFI is a near-infrared imager and grism spectrograph at
the ESO NTT 3.6m telescope. It is used to complement WFI observations in
near-infrared pass-bands. More specifically, it disposes of
$J$, $H$ and $K_{\rm s}$ filters 
that have some similarity to Glass $JHK$ (see Fig.~\ref{fig_filtri}). 
 
VEGAmag magnitudes are computed assuming the same as before, i.e.
that Vega apparent magnitudes are $0.03$.

\subsubsection{EMMI}
\label{sec_emmi}

EMMI is a visual imager and grism spectrograph also at
the NTT. The EIS project uses a set of its
wide filters, i.e.\ $B_{\rm w}$, $I_{\rm w}$, $R_{\rm w}$, 
and $V_{\rm w}$, together with the
$R$ filter. The pass-bands have been kindly provided by S.\ Arnouts. 
As before, VEGAmag magnitudes are computed assuming that Vega 
has $0.03$ mag in all filters.

\subsection{Washington system}
\label{sec_washington}

The Washington system, originally defined by G.\ Wallerstein and 
developed by Canterna (1976), has been more and more used
since Geisler (1996) defined a set of CCD standard fields.
In the present paper, we reproduce the Washington 
CCD system (filters $C$, $M$, $T_1$ and $T_2$) by adopting 
the transmission curves as revised by Bessell (2001). 

Following Geisler (1996), the Washington $T_1$ filter has been
frequently replaced by the Kron-Cousins filter $R$, which presents 
a similar transmission curve and is available in almost
all observatories. Additionally, Holtzman et al.\ (in preparation) 
suggest the use of $BVI$ colours together with Washington 
ones for breaking the age-metallicity degeneracy
of stellar populations in colour-colour 
diagrams. For these reasons, in all tables that deal
with the Washington $C M T_1 T_2$ system, we also insert the 
information for the Johnson-Cousins $BVRI$ filters.

Finally, following Geisler (private communication),
the zero-points should be well represented by a VEGAmag
system. We assume Vega has magnitude $0.03$ in all filters.

\subsection{Other planned systems}
\label{sec_futuresystems}

Forthcoming papers of this series will be dedicated to
\begin{itemize}
\item other traditional systems like Str\"omgren, 
Thuan-Gunn, Vilnius, etc.;
\item systems corresponding to some successful and 
recent/ongoing observational campaigns, like Hipparcos, 
MACHO, DENIS, 2MASS, and SDSS;
\item newly-proposed systems, like the ones for the future
astrometric mission GAIA.
\end{itemize}

Since the procedure for computing $BC_{S_\lambda}$ 
tables is relatively 
easy, provided the necessary information -- i.e. filter transmission 
curves and a zero-point definition that corresponds to one of the 
cases discussed in Sect.~\ref{sec_zeropoint} -- is given, we can provide
isochrone tables for any photometric systems upon request.

\section{Available stellar tracks and isochrones}
\label{sec_tracks}

The tables of bolometric corrections here described have been 
primarily constructed to be applied to the Padova database of
stellar evolutionary tracks and isochrones\footnote{Of course,
they can also be applied to any other set of stellar data in
the literature.}. These latter have been described in several 
previous papers, and a complete description of them is 
beyond the scope of this work.

In the following, our intention is just to briefly mention
the sets of isochrones which are the most useful for 
comparisons with observed photometric data, and to first 
mention some data that has not been published in precedence. 
A summary table of the available material is presented in 
Table~\ref{tab_isoc}. 

\begin{table*}
\caption{Stellar tracks used in the different set of isochrones.}
\label{tab_isoc}
\begin{tabular}{lll|lllll|lll}
	\hline\noalign{\smallskip}
\multicolumn{3}{l|}{Initial chemical comp.:} & 
\multicolumn{5}{l|}{Evolutionary tracks:} & 
\multicolumn{3}{l}{Isochrones:} \\
	\noalign{\smallskip}\cline{1-3} 
	\cline{4-8} \cline{9-11}\noalign{\smallskip} 
\multicolumn{3}{l|}{} & 
\multicolumn{3}{l|}{Mass ranges (in \Msun):\protect\footnotemark[2]} 
& \multicolumn{1}{l|}{TP-AGB} & convec- &
age range & filename in & basic \\
$Z$ & $Y$ & kind\protect\footnotemark[1] & $0.15$--$0.55$ & $0.6$--$7.0$ & 
\multicolumn{1}{l|}{$>7.0$} & \multicolumn{1}{l|}{evolution\protect\footnotemark[3]} & tion\protect\footnotemark[4] &
	$\log(t/{\rm yr})$ & database & reference\protect\footnotemark[2] \\
\noalign{\smallskip}\hline\noalign{\smallskip}
0.0    & 0.230 & S & --   & Ma01 & Ma01  & -- & O & $6.30 - 10.25$ & isoc\_z0.dat & Ma01 \\
0.0001 & 0.230 & S & Gi01 & Gi01 & Gi96  & G & O & $6.60 - 10.25$ & isoc\_z0001.dat & Gi01+Gi96 \\
0.0004 & 0.230 & S & Gi00 & Gi00 & Fa94a & G & O & $6.60 - 10.25$ & isoc\_z0004.dat & Gi00+Be94 \\
0.001  & 0.230 & S & Gi00 & Gi00 & --    & G & O & $7.80 - 10.25$ & isoc\_z001.dat & Gi00 \\
0.004  & 0.240 & S & Gi00 & Gi00 & Fa94b & G & O & $6.60 - 10.25$ & isoc\_z004.dat & Gi00+Be94  \\
0.008  & 0.250 & S & Gi00 & Gi00 & Fa94b & G & O & $6.60 - 10.25$ & isoc\_z008.dat & Gi00+Be94 \\
0.019  & 0.273 & S & Gi00 & Gi00 & Br93  & G & O & $6.60 - 10.25$ & isoc\_z019.dat & Gi00+Be94 \\
0.030  & 0.300 & S & Gi00 & Gi00 & --    & G & O & $7.80 - 10.25$ & isoc\_z030.dat & Gi00 \\
\noalign{\smallskip}\hline\noalign{\smallskip}
0.019  & 0.273 & S & Gi00 & Gi00 & --    & G & C & $7.80 - 10.25$ & isoc\_z019nov.dat & Gi00 \\
\noalign{\smallskip}\hline\noalign{\smallskip}
0.008  & 0.250 & S & Sa00 & Sa00 & Sa00  & G & O & $7.00 - 10.25$ & isoc\_z008s.dat & Sa00 \\
0.019  & 0.273 & S & Sa00 & Sa00 & Sa00  & G & O & $7.00 - 10.25$ & isoc\_z019s.dat & Sa00 \\
0.040  & 0.320 & S & Sa00 & Sa00 & Sa00  & G & O & $7.00 - 10.25$ & isoc\_z040s.dat & Sa00 \\
0.070  & 0.390 & S & Sa00 & Sa00 & Sa00  & G & O & $7.00 - 10.25$ & isoc\_z070s.dat & Sa00 \\
0.008  & 0.250 & A & Sa00 & Sa00 & Sa00  & G & O & $7.00 - 10.25$ & isoc\_z008a.dat & Sa00 \\
0.019  & 0.273 & A & Sa00 & Sa00 & Sa00  & G & O & $7.00 - 10.25$ & isoc\_z019a.dat & Sa00 \\
0.040  & 0.320 & A & Sa00 & Sa00 & Sa00  & G & O & $7.00 - 10.25$ & isoc\_z040a.dat & Sa00 \\
0.070  & 0.390 & A & Sa00 & Sa00 & Sa00  & G & O & $7.00 - 10.25$ & isoc\_z070a.dat & Sa00 \\
\noalign{\smallskip}\hline\noalign{\smallskip}
0.004  & 0.240 & S & Gi00 & Gi00 & --    & M & O & $7.80 - 10.25$ & isoc\_z004m.dat & MG01 \\
0.008  & 0.250 & S & Gi00 & Gi00 & --    & M & O & $7.80 - 10.25$ & isoc\_z008m.dat & MG01 \\
0.019  & 0.273 & S & Gi00 & Gi00 & --    & M & O & $7.80 - 10.25$ & isoc\_z019m.dat & MG01 \\
\noalign{\smallskip}\hline\noalign{\smallskip}
\end{tabular} \\
\\
$^1$
S indicates a solar-scaled distribution 
of metals, A indicates an $\alpha$-enhanced one. The adopted 
metal abundance ratios are specified in Salasnich et al.\ (2000).
\\
$^2$
References for tracks and isochrones: 
Be94 = Bertelli et al.\ (1994, A\&AS 106, 275); 
Br93 = Bressan et al.\ (1993, A\&AS 100, 647); 
Fa94a = Fagotto et al.\ (1994a, A\&AS 104, 365); 
Fa94b = Fagotto et al.\ (1994b, A\&AS 105, 29); 
Gi96 = Girardi et al.\ (1996, A\&AS 117, 113); 
Gi00 = Girardi et al.\ (2000, A\&AS 141, 371); 
Gi01 = Girardi (2001, unpublished);
Ma01 = Marigo et al.\ (2001, A\&A 371, 152);
MG01 = Marigo \& Girardi (2001, A\&A 377, 132);
Sa00 = Salasnich et al.\ (2000, A\&A 361, 1023).
\\
$^3$
G means a simple synthetic evolution as in 
Girardi \& Bertelli (1998), whereas M stands for more detailed 
calculations as in Marigo (2001; and references therein).
\\
$^4$
O means a model with overshooting
(see Gi00 for all references), whereas C corresponds to classical 
semi-convective models.
\end{table*}

\subsection{An extended set of isochrones}
\label{sec_basic}

Girardi et al.\ (2000) computed a set of low- and 
intermediate-mass stellar tracks which supersedes
those previously used in Bertelli et al.\ (1994) isochrones.
Additional models (unpublished) 
have been recently computed for initial chemical
composition $[Z=0.0001, Y=0.23]$.
Thus, we have created a new set of isochrones combining 
the latest low- and intermediate mass tracks (in the range 
$0.15 - 7.0$~\Msun) with the formerly
available massive ones. The full references are given in
Table~\ref{tab_isoc}. 

In all cases in which ``2000'' and ``1994'' tracks have been combined,
we find excellent agreement between the relevant quantities (lifetimes,
tracks in the HR diagram) at the transition mass of $7-8$~\Msun. 
This is explained considering
that the intermediate- and high-mass models share the 
same prescription for convection, and have interior opacities dominated 
by electron scattering (which has not been changed in the meanwhile). 
For the solar metallicity, we have combined tracks presenting 
slightly different initial metallicities -- 
$[Z=0.019, Y=0.273]$ in Girardi et al.\ (2000), 
and $[Z=0.020, Y=0.280]$ in Bressan et al.\ (1993) --
without finding any significant discontinuity.

Finally, to this extended set we have included the Marigo et al.\ 
(2001) isochrones for zero-metallicity stars.

\subsection{Overshooting vs.\ classic models}
\label{sec_overshoot}

For solar metallicity and in the mass range $0.15 - 7.0$~\Msun,
Girardi et al.\ (2000) presents an additional set of tracks and 
isochrones computed with the
classical semi-convective prescription for convective borders.
Then, the two $[Z=0.019, Y=0.273]$ sets are useful if one 
chooses to compare classical with overshooting models. 
The same kind of work is being extended to metallicities 
$[Z=0.004, Y=0.240]$ and $[Z=0.008, Y=0.250]$ (Barmina et al.\ 2002),
and will be included in the database when completed.

\subsection{Solar-scaled vs.\ $\alpha$-enhanced models}
\label{sec_alphaen}

Salasnich et al.\ (2000) presents new
models for 4 different metallicities 
($[Y=0.250, Z=0.008]$, $[Y=0.273, Z=0.019]$, $[Y=0.320, Z=0.040]$ 
and $[Y=0.390, Z=0.070]$), computed both with scaled-solar and
alpha-enhanced distributions of metals. The tracks cover the
mass range from $0.15$ to $20$~\Msun. They represent a valid 
alternative to the isochrones referred to in Sect.~\ref{sec_basic},
but do not cover the low-metallicity interval.
Anyway, it has been demonstrated by Salaris et al.\ (1993; see
also Salaris \& Weiss 1998; and VandenBerg 2000), 
that for low metallicities one may safely use scaled-solar
stellar models instead of $\alpha$-enhanced ones of same $Z$.

\subsection{Simple vs.\ more detailed TP-AGB models}
\label{sec_tpagb}

All the tracks and isochrone sets above mentioned, include the 
complete TP-AGB phase as computed with a simple synthetic algorithm
(Girardi \& Bertelli 1998). In parallel,
Marigo (2001, and references therein) has developed a much more
sophisticated code for synthetic TP-AGB evolution, which includes 
crucial processes such as the third dredge-up and hot-bottom burning.
Sets of complete TP-AGB tracks have been so far presented
for metallicities $[Y=0.240, Z=0.004]$, $[Y=0.250, Z=0.008]$, and
$[Y=0.273, Z=0.019]$ (Marigo 2001). A set of isochrones has been 
generated by combining these detailed TP-AGB tracks with the previous 
evolution from Girardi et al.\ (2000); they are
presented in the appendix of Marigo \& Girardi (2001).

These isochrones represent a useful alternative to the isochrones 
referred to in Sect.~\ref{sec_basic}, any time the 
TP-AGB population is under scrutiny -- for instance, when we have 
near-infrared photometry of objects above the RGB-tip. 
Marigo TP-AGB models are being extended to cover a larger
metallicity interval.

\section{Discussion and concluding remarks}
\label{sec_conclu}

\subsection{Summary}

This work is dedicated to the presentation 
of theoretical isochrones in several photometric systems.
It represents the continuation of a wide project of the Padova
group, started with Bertelli et al.\ (1994) and then carried on 
by Chiosi et al.\ (1997), and more recently by Salasnich et al.\ 
(2000). It starts describing the formalism for
converting synthetic stellar spectra into tables of bolometric
corrections (Sect.~\ref{sec_synphot}), in such a way that it 
can be easily applied to different photometric systems, and  
with the possibility of including extinction in a self-consistent way.

	\begin{figure*}
	\resizebox{\hsize}{!}{\includegraphics{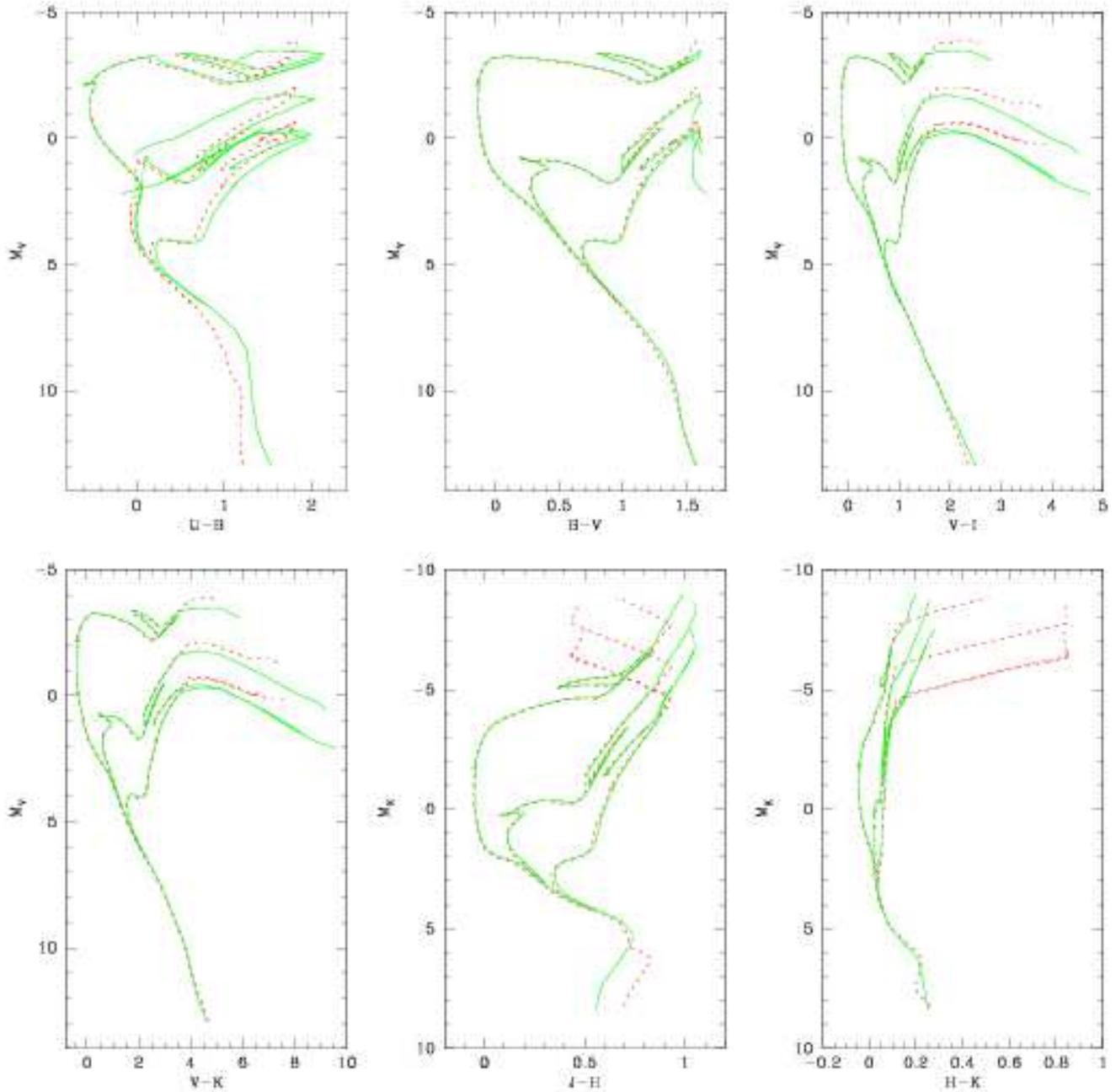}}
	\caption{Comparison of Girardi et al.\ (2000)
isochrones transformed to Johnson-Cousins-Glass magnitudes and 
colours by using either present 
relations (continuous lines) or Bertelli et al.\ (1994) ones
(dashed lines), in several CMDs.
The isochrones have solar metallicity ($Z=0.019$) and ages
$10^8$, $10^9$, and $10^{10}$~yr (from top to bottom).}
	\label{fig_cmdjohnoldnew}
	\end{figure*}

	\begin{figure*}
	\resizebox{\hsize}{!}{\includegraphics{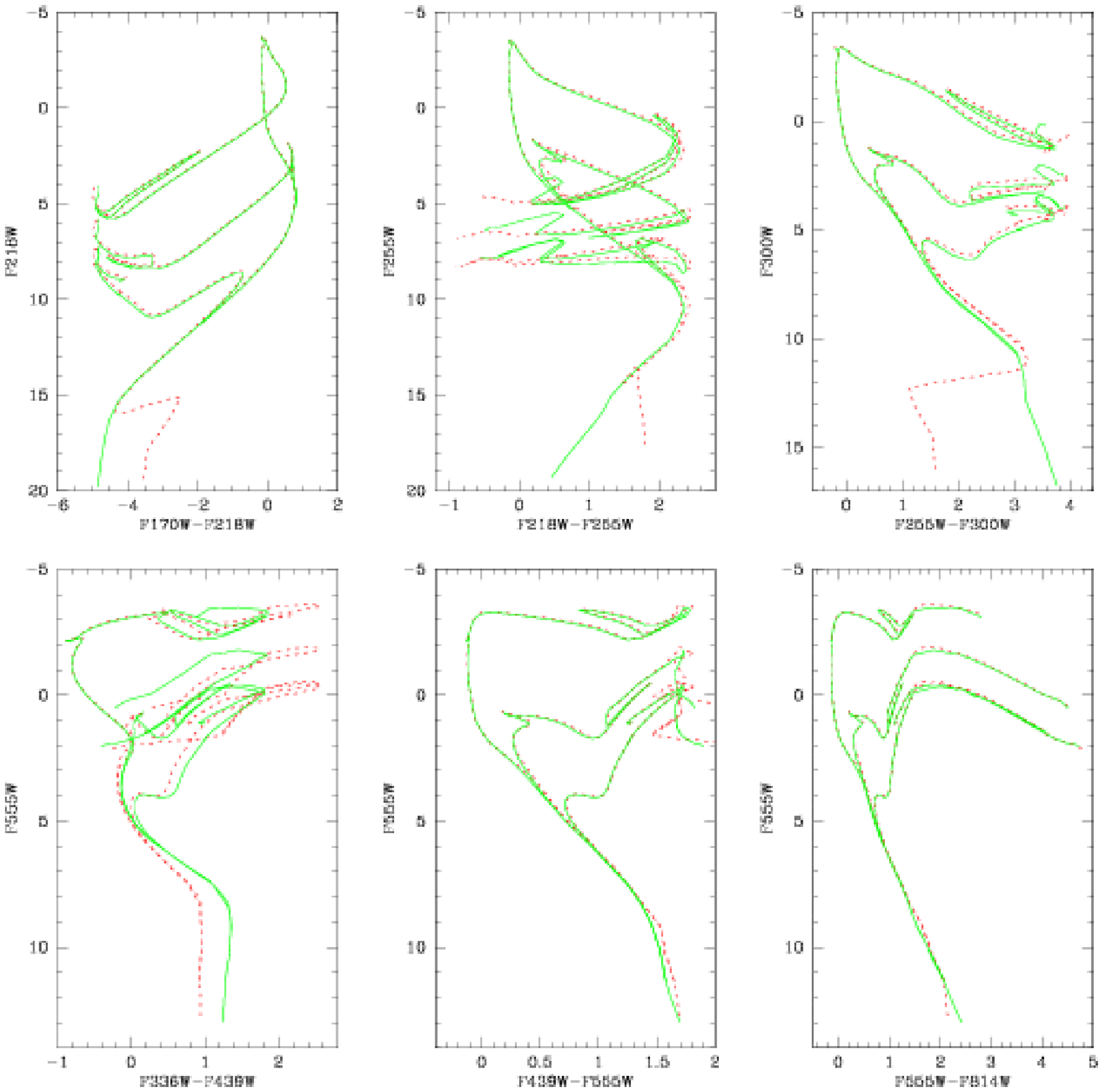}}
	\caption{Comparison of Girardi et al.\ (2000)
isochrones as transformed to the WFPC2 VEGAmag system
using either present relations (continuous lines) or 
Salasnich et al.\ (2000) ones (dashed lines). 
The isochrone ages and metallicities 
are the same as in Fig.~\protect\ref{fig_cmdjohnoldnew}.}
	\label{fig_cmdwfpc2oldnew}
	\end{figure*}

Then, we describe the assemblage of an updated library of stellar
spectra (Sect.~\ref{sec_spectra}). The library is quite
extended in \Teff\ and $\log g$, and includes the crucial dependence 
of spectral features on metallicity. Of course, it suffers from some
limitations:
very hot stars and M-giants are not included among synthetic spectra,
a situation which can be remediated by using blackbody and empirical
spectra, respectively. The strinkingly different spectra of
carbon stars will also have to be 
considered in the future (Marigo et al., in preparation). 
These are probably the points where the models can be most improved. 
Moreover, several problems may be affecting the ATLAS9 
synthetic spectra we are using to simulate the broad-band 
colours of most stars
(see Sect.~\ref{sec_kurucz}). Although such spectra
have been demonstrated to be suitable for 
synthetic photometry (mainly for the Johnson-Cousins-Glass 
system; e.g.\ Bessell et al.\ 1998), their accuracy
has still to be sistematically evaluated for stars of all
metallicities, temperatures and gravities. 
Anyway, we assume that they 
are good enough for simulating broad-band photometric systems
in the visual-infrared wavelength region, whereas 
expect that the results in narrow-band systems, and in the 
ultraviolet pass-bands, will be affected by more significant 
errors. 
Another important aspect of synthetic spectra is that they
are usually computed for scaled-solar chemical compositions, 
whereas the extension to peculiar and $\alpha$-enhanced mixtures 
would be of high interest. This latter problem will probably be 
alleviated in a near future, with the release of extensions 
to ATLAS9 spectra.

From the spectral library, we derive bolometric corrections for each 
pass-band mentioned in Sect.~\ref{sec_systems}, and apply them 
to the Padova isochrones. In practice, only in few
cases we present newly-constructed isochrones: the bulk of 
isochrone data is already described in previous papers by our group,
and
it is only the transformation from theoretical quantities to
absolute magnitudes that changes compared to past releases.
Sect.~\ref{sec_tracks} summarizes the 
basic characteristics of the different sets
of isochrones, indicating the full references. 

\subsection{Comparison with previous works}

It is important to illustrate the differences between the
present and previous transformations. The previous ones for 
$UBVRIJHK$ are fully described in Bertelli et al.\ (1994), and 
were adopted by Girardi et al.\ (2000), Salasnich et al.\ (2000)
and Marigo et al.\ (2001); for HST/WFPC2 photometry, they are the 
ones described in Salasnich et al.\ (2000). 

The situation for Johnson-Cousins-Glass is tentatively illustrated in 
Fig~\ref{fig_cmdjohnoldnew}, which compares a set of 
Girardi et al.\ (2000) isochrones, transformed according to both 
present (continuous lines) and Bertelli et al.\ (1994;
dashed lines) transformations. We point out that: 
	\begin{enumerate}
	\item
For stars hotter than $\Teff\sim4\,000$~K, present transformations 
are very similar to those of Bertelli et al.\ (1994).
The differences amount to just a few 
hundredths of a magnitude over most regions of the CMD, including 
the entire main sequence and subgiant branch, and most of the RGB.
They can be entirely attributed to the slightly different pass-bands 
and zero-points, and to the use of more recent ATLAS9 ``NOVER'' 
atmospheres instead of Kurucz (1993) ones.
	\item
A somewhat similar situation holds for dwarfs cooler 
than $\Teff\sim4\,000$~K (see bottom end of the main sequence in 
all panels, for $\mv\ga7$, and $\mk\ga5$).
The differences are generally small and can
be attributed to the change from Kurucz (1993) to 
Allard et al.\ (2000a) spectra. The exceptions are \ub\ and 
\jh\ colours, for which the differences between the two versions
become sizeable.
	\item
For giants cooler than $4\,000$~K, present transformations 
become very different. This can be noticed in the upper-right 
corner of all diagrams. M-giants corresponding to the 
RGB-tip and TP-AGB, are now seen to 
fade by some magnitudes in $V$, due to a sort of rapid increase 
of visual BCs at $\Teff\la3\,500$~K. This effect is caused by the
use of Fluks et al.\ (1994) spectra and their
\Teff\ vs.\ \vk\ scale. Such a bending of the RGB is indeed 
observed in CMDs of old metal-rich clusters 
(see e.g.\ Ortolani et al.\ 1990; and Rich et al.\ 1998), 
and seems to be better 
reproduced now than with previous transformations.
Other differences appear in all colours: the most remarkable 
are the excursion of M-giants of latest type towards much
bluer \ub, and the much smoother behaviour now obtained 
for \jh\ and \hk.
	\end{enumerate}

A quite similar situation holds for HST/WFPC2 photometry, 
as illustrated in Fig.~\ref{fig_cmdwfpc2oldnew}. This time,
we compare the same isochrones as transformed with present 
(continuous lines) and Salasnich et al.\ (2000; dashed lines) 
transformations. It is evident that the present transformations 
ensure a more continuous behaviour of the colours for all low-temperature
stars (both dwarfs and giants). 

From the plots at the top row of Fig.~\ref{fig_cmdwfpc2oldnew}, one can
also appreciate the unusual appearance of isochrones in CMDs 
that involve ultraviolet WFPC2 pass-bands: Notice for instance that
in F170W, F218W, F255W and F330W magnitudes, giants may be {\rm fainter} 
than turn-off stars. Isochrones in the F218W vs.\ F170W$-$F218W and 
F255W vs.\ F218W$-$F255W diagrams are even ``twisted'', because the 
\Teff\ vs.\ colour relations are not monotonic for
these filters. These effects are related to the presence of a 
red leak in the ultraviolet HST filters (for both the present WFPC2 
and the former FOC camera), and are extensively 
discussed by Yi et al.\ (1995) and Chiosi et al.\ (1997).

As a consequence of the great similarity between present and 
previous $UBVRIJHK$ and HST/WFPC2 transformations, 
for most colours and over a large portion of the HR diagram, 
most results derived from previous Padova isochrones 
are not expected to change. 
Exceptions may show up for works that are concerned with
the photometry of the reddest giants, with $\Teff\la3\,500$~K 
($\bv\ga1.5$), or that deal with low-mass main-sequence stars 
in the \ub\ and ultraviolet colours.

	\begin{figure*}
	\begin{minipage}{0.70\hsize}
	\resizebox{\hsize}{!}{\includegraphics{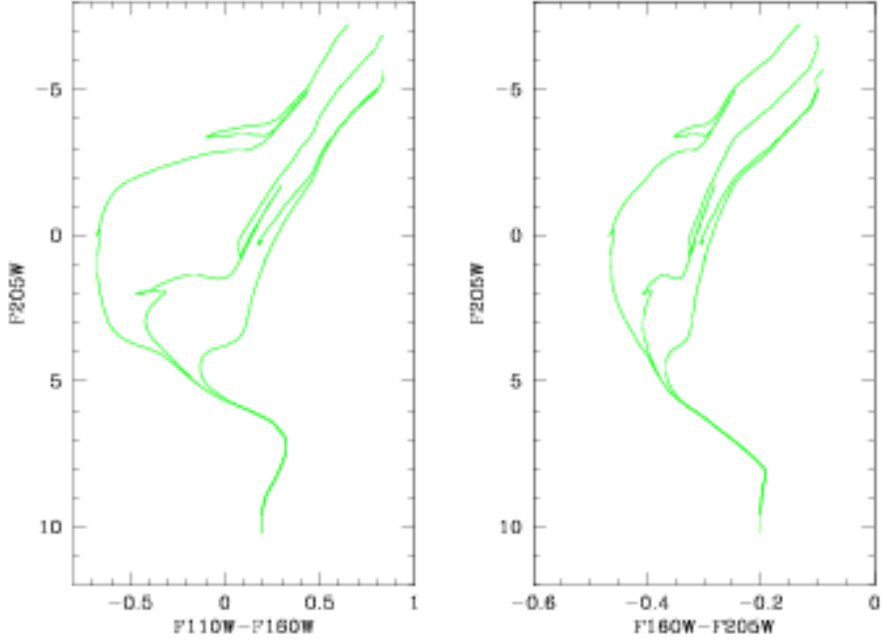}}
	\end{minipage}
	\hfill
	\begin{minipage}{0.28\hsize}
	\caption{Isochrones in the CMDs of NICMOS ABmag photometry. 
Ages and metallicities are the same as in 
Fig.~\protect\ref{fig_cmdjohnoldnew}.}
	\label{fig_cmdnicmos}
	\end{minipage}
	\end{figure*}

	\begin{figure*}
	\begin{minipage}{0.70\hsize}
	\resizebox{\hsize}{!}{\includegraphics{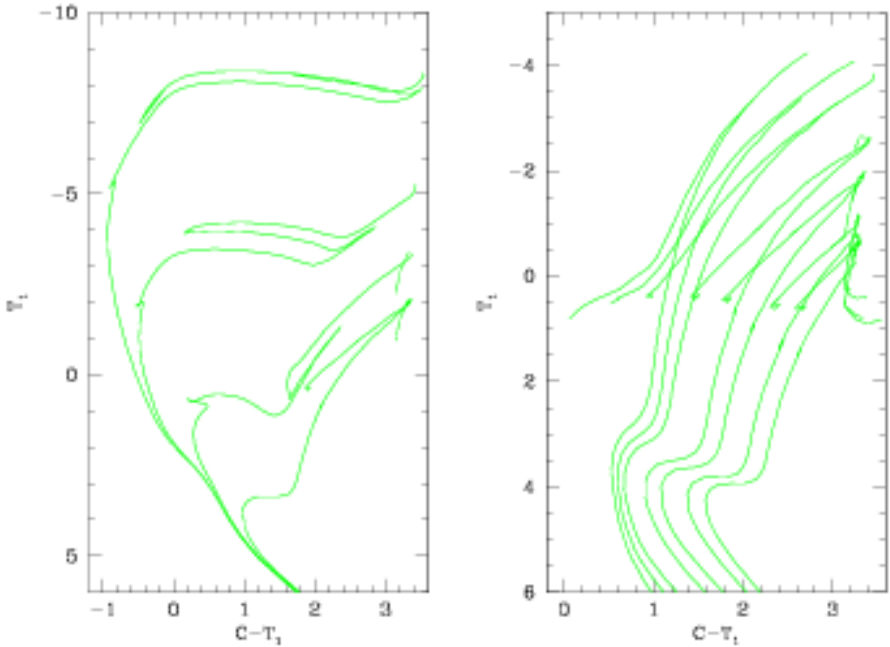}}
	\end{minipage}
	\hfill
	\begin{minipage}{0.28\hsize}
	\caption{Isochrones in the $T_1$ vs.\ $C-T_1$ plane
of Washington photometry. Left panel: from top to bottom,
a sequence of $Z=0.008$ isochrones with ages $10^7$, 
$10^8$, $10^9$, and $10^{10}$~yr. Right panel: from left to right, 
a sequence of 14~Gyr old isochrones with metallicities
$Z=0.0001$, $0.0004$, $0.001$, $0.004$, $0.008$, $0.019$,
and $0.030$. }
	\label{fig_cmdwashington}
	\end{minipage}
	\end{figure*}

\subsection{New results}

\begin{figure*}
	\resizebox{\hsize}{!}{\includegraphics{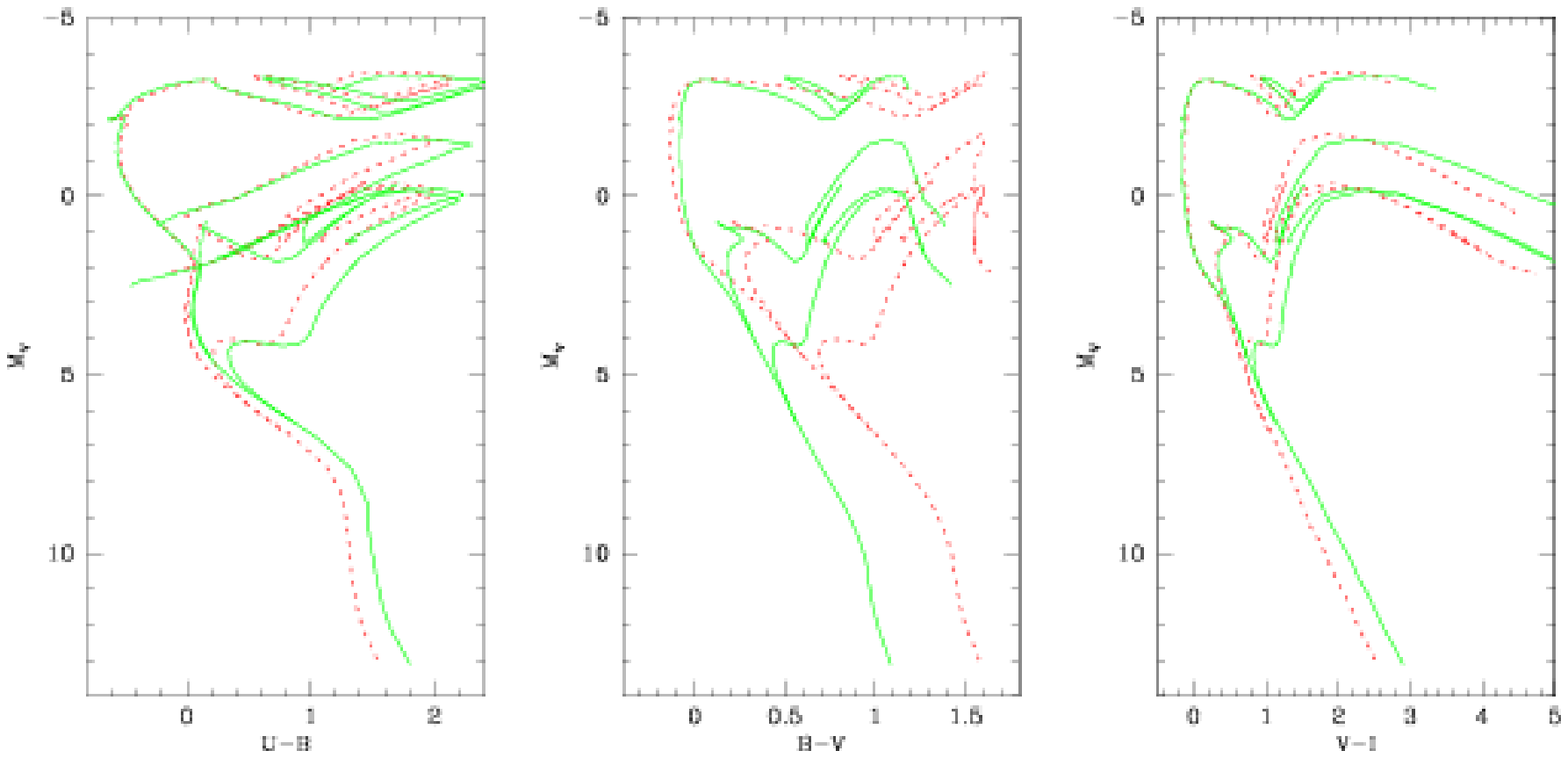}}
	\caption{Comparison between the same set of
$Z=0.019$ isochrones as seen in the $UBVI$ CMDs 
using either Johnson-Cousins (dashed lines) or WFI filters
(continuous lines). VEGAmag systems are used in both cases.
Ages are the same as in Fig.~\protect\ref{fig_cmdjohnoldnew}. }
	\label{fig_cmdeis}
	\end{figure*}

The greatest improvement of the present database is
in the presentation of Padova isochrones in several 
photometric systems for which they were not available so far -- 
including the case of brand-new systems. 
Three examples of theis kind are given 
in Figs.~\ref{fig_cmdnicmos}, \ref{fig_cmdwashington} 
and \ref{fig_cmdeis}. 

First, Fig.~\ref{fig_cmdnicmos} presents the isochrones in
NICMOS ABmag system. In a VEGAmag system, 
NICMOS isochrones would look similar to their equivalent 
Johnson-Cousins-Glass ones, shown in Fig.~\ref{fig_cmdjohnoldnew}.
In the ABmag system, however, they appear shifted to quite
different colour and magnitude intervals.

Figure.~\ref{fig_cmdwashington} illustrates how Padova 
isochrones look like in the $T_1$ vs.\ $C-T_1$ CMD of Washington 
photometry, both for varying age at constant metallicity 
(left panel), and for varying metallicity at constant age 
(right one). The striking feature in these plots is the excellent 
separation in metallicity offered by the $C-T_1$ colour, 
from the main sequence up to red giant phases. 
This feature, combined to the excellent throughput in the $C$ filter 
(Fig.~\ref{fig_filtri}), is among the 
advantages that make the Washington system a very competitive one 
if compared to Johnson-Cousins (see also Paltoglou \& Bell 1994, 
and Geisler \& Sarajedini 1999). 

Preliminary comparisons point to a good agreement 
between our Washington isochrones and real data for LMC 
fields from Bica et al.\ (1998). Just to mention an 
example, we notice that $C-T_1$ for giants 
``saturates'' at $\sim3.4$, both in the models and in 
the LMC data. 

An example of ``new'' photometric system is provided
by the WFI, which has broad-band filters very different 
from Johnson-Cousins ones. To illustrate the effect in
colours, Fig.~\ref{fig_cmdeis} shows exactly the same 
isochrones as seen in $BVI$ CMDs 
using either Johnson-Cousins or WFI filters, and applying
in both cases the VEGAmag definition of zero-points. 
The differences are striking. In particular, since the $BV$ 
WFI filters represent a wavelength baseline shorter than the
Johnson ones, they provide a more modest separation of stars 
in $\bv$ colour. It is evident from this plot that the normal 
Johnson-Cousins isochrones cannot be used to interpret WFI data that 
has been converted to VEGAmag or ABmag systems, as for most of
EIS data (e.g.\ Arnouts et al.\ 2001; Groenewegen et al.\ 
2002)\footnote{Notice, 
however, that part of the data released from EIS 
-- namely the pre-FLAMES survey -- is in fact converted
into a standard Johnson-Cousins system (see Momany et al.\ 2001).}. 

\subsection{Retrieval of electronic tables}

All the data here mentioned are available at the WWW site
\verb$http://pleiadi.pd.astro.it$. The database already includes a
very large number of files, and is expected to increase further
as we publish data for other photometric systems. Thus, it is 
hard to describe here both the structure of the database,
and the content of each file. Moreover, this kind of information
is probably useful just to whom actually accesses the database.
Thus, we opt to provide all the relevant information 
in \verb$readme.txt$ files inserted in the database. 

To the general reader, suffice it to briefly mention the kind of
data which is available: 
	\begin{itemize}
	\item Tables of bolometric corrections for each metallicity: 
they contain the quantities $BC_{S_\lambda}$ for each filter, and
as a function of stellar \Teff\ and $\log g$. Metallicities available
are $\feh=-2.0$, $-1.5$, $-1.0$, $-0.5$, $0$, $+0.5$. The values
of  \Teff\ and $\log g$ are not exactly the same for all 
metallicities, but correspond quite well to the regions indicated 
in Fig.~\ref{fig_griglia}. 
	\item Tables of isochrones: they include all metallicities
and cases indicated in Sect.~\ref{sec_tracks}. The various
photometric systems are separated in different directories. 
For each isochrone file, the information and structure are the
same as already presented in Girardi et al.\ (2000), Salasnich et 
al.\ (2000), and Marigo \& Girardi (2001), with the obvious 
difference that instead of $UBVRIJHK$ absolute magnitudes, in each
file we tabulate the absolute magnitudes for the photometric
system under consideration. 
	\item Tables of integrated colours of single-burst stellar 
populations: a table of this kind is present for each 
isochrone table. They provide integrated magnitudes in each 
pass-band, as a function of age.
	\end{itemize}

\begin{acknowledgements}
L.G.\ thanks the many people who 
helped by providing filter transmission curves
and zero-points information (in particular E.\ Bica, 
D.\ Geisler, J.\ Holtzman,  D.\ Figer, E.\ Grebel, M.\ Gregg, 
M.\ Rich, S.\ Arnouts, and L.\ da Costa). Particularly appreciated 
are the availability (R.\ Kurucz, F.\ Allard) and help with 
(I.\ Baraffe) on-line spectral data, the useful comments by B.\ Plez
regarding cool giants, and the many useful remarks by R.\ Bell and 
M.S.\ Bessell, which greatly helped to improve this paper.
Also acknowledged are those who 
kindly pointed out some mistakes in our preliminar releases 
of data. L.G.\ acknowledges a stay at MPA funded by the European TMR
grant ERBFMRXCT 960086. 
This work was partially funded by the Italian MURST.

\end{acknowledgements}


\section*{References}

\begin{description}
\item Allard F., Hauschildt P.H., Alexander D.R., \& Starrfield S.,
	1997, ARA\&A 35, 137  
\item Allard F., Hauschildt P.H., Alexander D.R., Tamanai A., \& 
	Ferguson J.W., 2000a, 
	in  proceed. of ``From giant planets to cool stars'',
	ASP Conf.\ Series v. 212, (eds.) C.A. Griffith \& M.S. Marley,
	p.\ 127 
\item Allard F., Hauschildt P.H., \& Schwenke D., 2000b, ApJ 540, 1005
\item Allard F., Hauschildt P.H., Alexander D.R., Tamanai A., \&
	Schweitzer A., 2001, ApJ 556, 357 
\item Allende Prieto C., Barklem P.S., Asplund M., \& Ruiz Cobo B.,
	2001, ApJ 558, 830
\item Allende Prieto C., Asplund M., Garcia L\'opez R.J., \& Lambert D.L.,
	2002, ApJ 567, 544
\item Alonso A., Arribas S., \& Mart\'\i nez-Roger C., 1998, A\&AS 131, 209
\item Alonso A., Arribas S., \& Mart\'\i nez-Roger C., 1999a, A\&AS 139, 335
\item Alonso A., Arribas S., \& Mart\'\i nez-Roger C., 1999b, A\&AS 140, 261
\item Alvarez R., Lan\c con A., Plez B., \& Wood P.R., 2000, A\&A 353, 322
\item Alvarez R., \& Plez B., 1998, A\&A 330, 1109
\item Arnouts S., Vandame B., Benoist C., et al., 2001, A\&A 379, 740
\item Asplund M., Ludwig H.-G., Nordlund A., \& Stein R.F., 2000, A\&A 359, 669
\item Baggett S., \& Gonzaga S., 1998, ISR WFPC2 98-03
\item Bahcall J.N., Pinsonneault M.H., \& Wasserburg G.J., 1995, 
	Rev.\ Mod.\ Phys.\ 67, n.4, 781 
\item Barklem P.S., Piskunov N., \& O'Mara B., 2000a, A\&A 355, 5
\item Barklem P.S., Piskunov N., \& O'Mara B., 2000b, A\&A 363, 1091
\item Barmina R., Girardi L., \& Chiosi C., 2002, A\&A 385, 847
\item Bell R.A., Paltoglou G., \& Tripicco M.J., 1994, MNRAS 268, 771
\item Bell R.A., Balachandran S.C., \& Bautista M., 2001, ApJ 546, L65
\item Bertelli G., Bressan A., Chiosi C., Fagotto F., \& Nasi E., 1994, 
	A\&AS 106, 275
\item Bessell M.S., 1979, PASP 91, 589
\item Bessell M.S., 1990, PASP 102, 1181
\item Bessell M.S., 2001, PASP 113, 66
\item Bessell M.S., \& Brett J.M., 1988, PASP 100, 1134
\item Bessell M.S., Castelli F., \& Plez B., 1998, A\&A 333, 231
\item Bica E., Geisler D., Dottori H., et al., 1998, AJ 116, 723
\item Bohlin R.C., Holm A.V., Harris A.W., \& Gry C., 1990, ApJS 73, 413
\item Bressan A., Fagotto F., Bertelli G., \& Chiosi C., 1993, 
	A\&AS 100, 647
\item Buser R., \& Kurucz R., 1978, A\&A 70, 555 
\item Canterna R., 1976, AJ 81, 228 
\item Castelli F., 1999, A\&A 281, 817
\item Castelli F., \& Kurucz R.L., 1994, A\&A 281, 817
\item Castelli F., \& Kurucz R.L., 2001, A\&A 372, 260
\item Castelli F., Gratton R.G., \& Kurucz R.L., 1997, A\&A 318, 841
\item Chabrier G., Baraffe I., Allard F., \& Hauschildt P.H., 2000,
	ApJ 542, 464
\item Chiosi C., Vallenari A., \& Bressan A., 1997, A\&AS 121, 301
\item Ciardi D.R., van Belle G.T., Thompson R.R., Akeson R.L., \& 
	Lada E.A., 2000, AAS 197, 4503
\item Code A.D., Bless R.C., Davis J., \& Brown R.H., 1976, ApJ, 203, 417
\item Colina L., Bohlin R., \& Castelli F., 1996, 
	Instrument Science Report CAL/SCS-008
\item da Costa L., 2000, in From Extrasolar Planets to Cosmology: 
	The VLT Opening Symposium, (eds.) J.\ Bergeron and A.\ Renzini,
	Springer-Verlag, Berlin, p.\ 192.
\item Edvardsson B., \& Bell R.A., 1989, MNRAS 238, 1121
\item Fagotto F., Bressan A., Bertelli G., \& Chiosi C., 1994a, A\&AS 104, 365
\item Fagotto F., Bressan A., Bertelli G., \& Chiosi C., 1994b, A\&AS 105, 29
\item Flower P.J., 1997, ApJ 469, 355
\item Fluks M.A, Plez B., The P.S., et al., 1994, A\&AS 105, 311
\item Fukugita M., Ichikawa T., Gunn J.E., et al., 1996, AJ 111, 1748
\item Geisler D., 1996, AJ 111, 480
\item Geisler D., \& Sarajedini A., 1999, AJ 117, 308
\item Girardi L., \& Bertelli G., 1998, MNRAS 300, 533
\item Girardi L., Bressan A., Chiosi C., Bertelli G., \& Nasi E., 1996,
	A\&AS 117, 113
\item Girardi L., Bressan A., Bertelli G., \& Chiosi C., 2000, A\&AS  141, 371
\item Grebel E.K., \& Roberts W.J., 1995, A\&AS 109, 293
\item Groenewegen M.A.T., Girardi L., Hatziminaoglou E., 
	et al., 2002, A\&A submitted.
\item Gunn J.E., \& Stryker L.L., 1983, ApJS 52, 121
\item Hayes D.S., 1985, Calibration of fundamental stellar quantities, 
  	IAU Symposium 111, ed.\ D.S.\ Hayes, L.E.\ Pasinetti and A.G.D.\ 
	Philip (Dordrecht, Reidel), p.\ 225
\item Hayes D.S., \& Lathan D.W., 1975, ApJ 197, 593
\item Hauschildt P.H., Allard F., Ferguson J., Baron E., \& Alexander D.R., 
	1999, ApJ 525, 871 
\item Holtzman J.A., Burrows C.J., Casertano S., et al., 1995, PASP 107, 1065
\item Houdashelt M.L., Bell R.A., Sweigart A.V., \& Wing R.F., 2000a, 
	AJ 119, 1424
\item Houdashelt M.L., Bell R.A., \& Sweigart A.V., 2000b, AJ 119, 1448
\item Kurucz R.L., 1993, in  IAU
     	Symp. 149: The Stellar Populations of Galaxies, 
	eds.\ B.\ Barbuy, A.\ Renzini, Dordrecht, Kluwer, p.\ 225
\item Kurucz R.L., 1995, in Astrophysical Applications of Powerful 
	New Databases. Joint Discussion No. 16 of the 22nd
	IAU General Assembly,  ASP Conference Series, V. 78,
	(eds.) S.J. Adelman, and W.L. Wiese, publisher: Astronomical
	Society of the Pacific, San Francisco, California, p.\ 205
\item Landolt A.U., 1992, AJ 104, 340
\item Leggett S.K., Allard F., Dahn C., et al., 2000, ApJ 535, 965
\item Leggett S.K., Allard F., Geballe T.R., Hauschildt P.H., \& 
	Schweitzer A., 2001, ApJ 548, 908
\item Lejeune T., Cuisinier F., \& Buser R., 1997, A\&AS 125, 229 
\item Lejeune T., Cuisinier F., \& Buser R., 1998, A\&AS 130, 65 
\item Lupton R.H., Gunn J.E., \& Szalay A.S., 1999, AJ 118, 1406 
\item Marigo P., 2001, A\&A 370, 194
\item Marigo P., \& Girardi L., 2001, A\&A 377, 132
\item Marigo P., Girardi L., Chiosi C., \& Wood P.R., 2001, A\&A 371, 152
\item Momany Y., Vandame B., Zaggia S., et al., 2001, A\&A 379, 436
\item Neckel H., \& Labs D., 1984, Sol. Phys, 90, 205
\item Oke J.B., 1964, ApJ 140, 689
\item Oke J.B., \& Gunn J.E., 1983, ApJ 266, 713
\item Ortolani S., Barbuy B., \& Bica E., 1990, 263, 362
\item Paltoglou G., \& Bell R.A., 1994, MNRAS 268, 793
\item Plez B., 1999, in Asymptotic Giant Branch Stars, IAU Symp.\ 191, 
	(eds.) T. Le Bertre, A. Lebre, and C. Waelkens, p. 75
\item Renzini A., \& da Costa L., 1997, The Messenger 87, 23
\item Rich R.M., Ortolani S., Bica E., \& Barbuy B., 1998, AJ 116, 1295
\item Ridgway S.T., Joyce R.R., White N.M., \& Wing R.F., 1980, ApJ 235, 126
\item Salaris M., Chieffi A., \& Straniero O., 1993, ApJ 414, 580
\item Salaris M., \& Weiss A., 1998, A\&A 335, 943
\item Salasnich B., Girardi L., Weiss A., \& Chiosi C., 2000, A\&A 361, 1023
\item Schmidt-Kaler T., 1982, in: Landolt-B\"ornstein, Neue Serie VI/2b. 
	Springer-Verlag, Berlin, 453-455, pp.\ 15-18
\item Straizys V., \& Zdanavicius K., 1965, Bull.\ Vilnius Obs.\ 14, 1
\item Thuan T.X., \& Gunn J.E., 1976, PASP 88, 543
\item Tsuji T., Ohnaka K., \& Aoki W., 1996, A\&A 305, 1
\item Tsuji T., Ohnaka K., \& Aoki W., 1999, ApJ 520, 119
\item VandenBerg D.A., 2000, ApJS 129, 315
\item Yi S., Demarque P., \& Oemler Jr.~A., 1995, PASP 107, 273
\item Worthey G., 1994, ApJS 95, 107
\end{description}  

\end{document}